\documentclass[12pt]{article}
\usepackage{graphics}

\newcommand{\pubnum}{429}
\newcommand{\data}{June 1999}
\newcommand{\text}[1]{\mbox{#1}}

\newcommand{\ba}{\begin{array}} 
\newcommand{\ea}{\end{array}} 

\newcommand{\sq}{\ensuremath{\tilde q}} 
\newcommand{\su}{\ensuremath{\tilde u}} 
\newcommand{\sd}{\ensuremath{\tilde d}} 
\newcommand{\stopp}{\ensuremath{\tilde t}}

\newcommand{\cplus}{\ensuremath{\chi^+}}

\newcommand{\pl}{P_L} 
\newcommand{\pr}{P_R}

\newcommand{\gsim}{\mbox{ \raisebox{-4pt}{${\stackrel{\textstyle >}{\sim}}$} }} 
\newcommand{\lsim}{\mbox{ \raisebox{-4pt}{${\stackrel{\textstyle <}{\sim}}$} }}

\newcommand{\limit}[2]{{\mbox{ 
                        $\longrightarrow$\hskip 0cm\kern-.6cm 
                        \vskip -.25cm 
                        ${\scriptscriptstyle #1\rightarrow #2}$} 
                     }}

\newcommand{\tb}{\ensuremath{\tan\beta}} 
\newcommand{\tbs}{\ensuremath{\tan^2\!\beta}} 
\newcommand{\ctb}{\ensuremath{\cot\beta}} 
\newcommand{\ctbs}{\ensuremath{\cot^2\!\beta}} 
\newcommand{\sbt}{\ensuremath{\sin\beta}} 
 
\newcommand{\cbt}{\ensuremath{\cos\beta}}

\newcommand{\sa}{\ensuremath{\sin\alpha}} 
 
\newcommand{\ca}{\ensuremath{\cos\alpha}}

\newcommand{\mb}{\ensuremath{m_b}} 
 
\newcommand{\mt}{\ensuremath{m_t}} 
\newcommand{\mts}{\ensuremath{m_t^2}}

\newcommand{\mw}{\ensuremath{M_W}} 
\newcommand{\mws}{\ensuremath{M^2_W}}

\newcommand{\mHp}{\ensuremath{M_{H^\pm}}} 
\newcommand{\mA}{\ensuremath{M_{A^0}}}

\newcommand{\mi}{\ensuremath{M_i}}

\newcommand{\mg}{\ensuremath{m_{\tilde{g}}}}

\newcommand{\msbo}{\ensuremath{m_{\tilde{b}_1}}}

\newcommand{\msto}{\ensuremath{m_{\tilde{t}_1}}}

\newcommand{\tch}{{t\rightarrow c\,h}} 

\renewcommand{\pl}{P_L} 
\renewcommand{\pr}{P_R} 
\renewcommand{\tb}{{\tan\beta}}

\newcommand{\mc}{\ensuremath{m_c}}
\newcommand{\mj}{\ensuremath{M_j}}

\newcommand{\Amaidc}{\ensuremath{A_{-a i}^{(d,c)}}}
\newcommand{\Apaidc}{\ensuremath{A_{+a i}^{(d,c)}}}
\newcommand{\Amaidt}{\ensuremath{A_{-a i}^{(d,t)}}}
\newcommand{\Apaidt}{\ensuremath{A_{+a i}^{(d,t)}}}
\newcommand{\Ambidc}{\ensuremath{A_{-b i}^{(d,c)}}}
\newcommand{\Apbidc}{\ensuremath{A_{+b i}^{(d,c)}}}

\newcommand{\Amajdc}{\ensuremath{A_{-a j}^{(d,c)}}}
\newcommand{\Apajdc}{\ensuremath{A_{+a j}^{(d,c)}}}

\newcommand{\Amaidu}{\ensuremath{A_{-a i}^{(d,u)}}}
\newcommand{\Apaidu}{\ensuremath{A_{+a i}^{(d,u)}}}

\newcommand{\msdal}{\ensuremath{m_{\tilde d_a}}}
\newcommand{\msdbe}{\ensuremath{m_{\tilde d_b}}}

\newcommand{\msqo}{\ensuremath{m_{\tilde q_1}}}
\newcommand{\msqt}{\ensuremath{m_{\tilde q_2}}}
\newcommand{\msqs}{\ensuremath{m_{\tilde q_6}}}

\newcommand{\mda}{\ensuremath{m_{d}}}
\newcommand{\mdas}{\ensuremath{m_{d}^2}}

\newcommand{\md}{\ensuremath{m_d}}

\newcommand{\kz}{\ensuremath{K^{0d}_{s}}}

\newcommand{\kpcl}{\ensuremath{K^{+cd}_{sL}}}
\newcommand{\kpcr}{\ensuremath{K^{+cd}_{sR}}}
\newcommand{\kptl}{\ensuremath{K^{+td}_{sL}}}
\newcommand{\kptr}{\ensuremath{K^{+td}_{sR}}}

\newcommand{\kpclj}{\ensuremath{K^{+cd}_{vL}}}
\newcommand{\kpcrj}{\ensuremath{K^{+cd}_{vR}}}

\begin{document}

\thispagestyle{empty}

\hbox{ 
    \vrule height0pt width5in 
    \vbox{\hbox{\rm  
     UAB-FT-\pubnum 
   }\break
 \hbox{KA-TP-6-99\hfill}
   \break\hbox{hep-ph/9906268\hfill}  
 \break\hbox{\data\hfill} 
   \hrule height.1cm width0pt} 
   } \vspace{3mm}

\begin{center}
{\large \textbf{FCNC top quark decays in the MSSM: a door to SUSY physics in
    high luminosity colliders?}}

\bigskip

{\large Jaume Guasch}\renewcommand{\thefootnote}{*}\footnote{%
Electronic address: guasch@itp.uni-karlsruhe.de}\renewcommand{\thefootnote}{%
\dag}\footnote{%
On leave from Grup de F{\'\i}sica Te{\`o}rica and Institut de F{\'\i}sica
d'Altes Energies, Universitat Aut{\`o}noma de Barcelona, E-08193, Bellaterra
(Barcelona) Catalonia, Spain.} \vskip3mm

\textsl{Institut f{\"u}r Theoretische Physik \\[0pt]
Universit{\"a}t Karlsruhe, D-76128 Karlsruhe, Germany}

\vskip 8mm

{\large Joan Sol{\`a}}\renewcommand{\thefootnote}{\ddag}\footnote{%
Electronic address: sola@ifae.es}\setcounter{footnote}{0} \vskip3mm

\textsl{Grup de F{\'\i}sica Te{\`o}rica and Institut de F{\'\i}sica d'Altes
Energies, Universitat Aut{\`o}noma de Barcelona, E-08193, Bellaterra
(Barcelona) Catalonia, Spain}

\vspace{1cm}
\end{center}

\vspace{0.3cm}

\begin{center}
\textbf{ABSTRACT}
\end{center}

\begin{quotation}
\noindent We study the FCNC top quark decays $t\rightarrow c\,h$ in the
framework of the MSSM, where $h\equiv h^{0},H^{0},A^{0}$ is any of the
supersymmetric neutral Higgs bosons. We include the leading set 
of SUSY-QCD and SUSY electroweak contributions. While the FCNC top
quark decay into the SM Higgs boson has such a negligible rate that will not
be accessible to any presently conceivable accelerator, we find that there
is a chance that the potential rates in the MSSM can be measured at the high
luminosity colliders round the corner, especially at the LHC and possibly at
a future LC, but we deem it difficult at the upgraded Tevatron. In view of
the large SUSY-QCD effects that we find in the Higgs channels, and due to
some discrepancies in the literature, we have revisited the FCNC top quark
decay into gluon, $t\rightarrow c\,g$, in our framework. 
We confirm that the possibility of sizeable rates does
not necessarily require a general pattern of gluino-mediated FCNC
interactions affecting both the LH and the RH sfermion sectors -- the LH one
being sufficient. However, given the present bounds on sparticle masses, 
the gluon channel turns out to lie just below the expected
experimental sensibility, so our general conclusion is that the Higgs
channels $t\rightarrow c\,h$ ({especially the one for the light
CP-even Higgs}) have the largest potential top quark FCNC rates in the MSSM,
namely of order $10^{-4}$.
\end{quotation}

\newpage

\section{Introduction}

\label{sec:intro}

The study of the virtual effects in top quark decays into Higgs bosons could
be the clue to physics beyond the Standard Model (SM). This fact has already
been demonstrated for top quark decays into charged Higgs bosons, $%
t\rightarrow \,H^{+}b$\,\cite{CGGJS}, within the context of the
Minimal Supersymmetric Standard Model (MSSM)\cite{MSSM}\footnote{%
The corresponding study for general two-Higgs-doublet models ($2$HDM's) is
also available in Ref.\,\cite{CGHS}.}. Indeed, the potential
existence of large quantum effects induced by supersymmetric particles in
certain regions of parameter space may lead to highly significant changes in
the partial width of that decay, and this feature could have a serious
impact on the Higgs searches at the Tevatron\,\cite{GSPL1,Fermilab}%
. A situation which is in contrast to the SM decay of the top quark, $%
t\rightarrow W^{+}\,b$, where the SUSY quantum effects are in general much
more modest\,\cite{GJHS}.

Similarly, one may expect that the top quark decays into the neutral Higgs
bosons of the MSSM may undergo relevant enhancements. Notice, however, that
in contradistinction to the charged current decays mentioned above, the loop
contributions are in this case the lowest order effects as the neutral Higgs
decays of the top quark are mediated by Flavor Changing Neutral Currents
(FCNC). Therefore, since the FCNC processes are rather suppressed the
possibility of MSSM enhancements should be very welcome, especially for the
physics program at the LHC and perhaps also at a future linear collider LC. 
In the LHC, for
example, the production of top quark pairs will be very high: $\sigma (t%
\overline{t})=800\;pb$ -- {roughly two orders of magnitude larger than that
of the Tevatron II at }$\sqrt{s}=2\,TeV$. In the so-called low-luminosity
phase ($10^{33}\,cm^{-2}s^{-1}$) of the LHC one expects about one $t\,\bar{t}
$-pair per second, that is to say of the order of ten million $t\,\bar{t}$%
-pairs per year \cite{Gianotti}. And this number will be augmented by one
order of magnitude in the high-luminosity phase ($10^{34}\,cm^{-2}s^{-1}$).
As for a future $e^{+}e^{-}$ linear collider running at e.g. $\sqrt{s}%
=500\;GeV$, one has a smaller cross-section $\sigma (t\bar{t})=650\;fb$ but
a higher luminosity factor ranging from $5\times 10^{33}\,cm^{-2}s^{-1}$ to $%
5\times 10^{34}\,cm^{-2}s^{-1}$ \cite{Miller} and of course a much cleaner
environment. {One thus expects that both the LHC and the LC will initially
deliver datasets of order }$10fb^{-1}/${year increasing to several }$%
100\,fb^{-1}/${year  in the high-luminosity phase. On the other hand, at
the Tevatron II during the highest luminosity era (TeV}$33${) one expects
typical datasets of }$30fb^{-1}${. It follows that if the branching ratios
of the FCNC decays are augmented by extra contributions beyond the SM, one
should be able to collect enough statistics (perhaps some few hundred to few
thousand events) from the combined output of these machines enabling us to
perform an efficient study of these rare decays.}

We should immediately point out that, within the strict context of the SM,
the possibility of detecting FCNC decays of the top quark is essentially
hopeless. In particular, it has recently been recognized that the FCNC rate
of the top quark decay into the SM Higgs boson ($t\rightarrow c\,H_{SM}$ )
is much smaller {\cite{Mele,Erratum}} than originally thought\,\cite
{GEilam}: 
It turns out that $BR(t\rightarrow c\,H_{SM})=1\cdot 10^{-13}-4\cdot
10^{-15}$ $(M_{Z}\leq M_{H}\leq 2\;M_{W})$\,\cite{Mele}, which
means that it is far out of the range to be covered by any presently
conceivable high luminosity machine. On the other hand, the situation with
the FCNC decays of the top quark into gauge bosons ($t\rightarrow c\,V$;$%
\;V\equiv \gamma ,Z,g$) is not much more promising in the SM, the branching
ratios being at most of order $10^{-12}$ for the photon, slightly above $%
10^{-13}$ for the $Z$-boson, and at most $10^{-10}$ for the gluon channel 
\cite{GEilam,FCNCSM}. Thus the highest SM rate, namely that of $t\rightarrow
c\,g$, is still $5$ ({resp. }$7$) orders of magnitude below the
feasible experimental possibilities at the LHC (resp. Tevatron II). Clearly,
detection of FCNC decays of the top quark at visible levels (viz. $%
\,BR(t\rightarrow c\,X)\gsim 10^{-5}-10^{-4}$) by any of the future high
luminosity colliders would be instant evidence of new physics!

Therefore, one may judiciously ask whether extra virtual effects beyond the
SM can help to bring the top quark FCNC decay ratios to within observable
levels. For example, for the $t\rightarrow c\,V$ decays one finds, within the
non-supersymmetric $2$HDM's, that there can be significant enhancements{\cite
{GEilam,FCNC2HDM} which, however, turn out to be insufficient. A similar situation
occurs in the more interesting case of the MSSM, where again in spite of the
potential enhancements the electroweak gauge boson channels fall short to be
detected\,}\cite{Yang}{-\cite{Divitiis}. Only the FCNC top quark
decay in the gluon channel could be fairly sensitive to the SUSY
corrections in non-negligible regions of the MSSM parameter space~\cite{Yang}%
-\cite{Divitiis}, a fact which we wish to revisit within our framework in
order to compare with the more exceptional possibilities offered by the
Higgs channels -- on which we will mainly concentrate\footnote{For recent
  studies on FCNC top quark decays beyond the MSSM, see Ref.\,\cite{FCNCRp} and
  references therein.}.}

These Higgs channels comprise the FCNC top quark decays {into the two
CP-even (``scalar'') states and the CP-odd (``pseudoscalar'') state of the
Higgs sector of the MSSM} \cite{Hunter}, 
\begin{equation}
t\rightarrow c\,h\ \ \ \ (h=h^{0},H^{0},A^{0}).  \label{tch}
\end{equation}
{Worth emphasizing} is the fact that in the MSSM (in contrast to
the SM or the unconstrained $2${HDM) at least one of these decays (viz. }$%
t\rightarrow c\,h^{0}${) is \textsl{always} possible, for in the MSSM there
is an upper bound on the mass of the lightest CP-even Higgs boson, }$%
M_{h^{0}}\lsim135\,GeV${\cite{Higgsloop1}-\cite{HollWeig},
which is below the top quark mass. Moreover, for  a sufficiently light
pseudoscalar mass, }$M_{A^{0}}<m_{t},$ {all three decays (\ref{tch}) are in
principle possible, if the SUSY masses are not that high so as to induce too
large positive corrections to }$M_{H^{0}}$.

As it is also the case with their charged-current counterpart mentioned in
the beginning, these decays could be greatly enhanced in wide regions of the
parameter space. Although some work already exists in the literature on FCNC
decays of the top quark into Higgs bosons within the framework of the
MSSM\,\cite{tch}, 
we feel that it is still rather incomplete because it fails to include some
of the most significant contributions and it does not make use of 
the one-loop Higgs mass relations 
of the MSSM \cite{Higgsloop1}-\cite{HollWeig}. 
These relations
play an essential role in 
correlating the various quantum effects for the different channels (\ref{tch})
and are fundamental in establishing the
aforementioned upper bound on the mass of the lightest supersymmetric 
Higgs particle.
Therefore, our purpose is to go beyond those
preliminary calculations and show, from a more rigorous and systematic
treatment of the different kinds of quantum effects and of the MSSM parameter space,
that the FCNC width of the top quark could in fact reach the experimentally visible
level at the high-luminosity colliders\,\cite{Prelim}. 

The paper is organized as follows. In Sect. 2 we give the SUSY Lagrangian
interactions relevant for the FCNC decays of the top quark. In Sect. 3 we
report on the SUSY electroweak (SUSY-EW) and supersymmetric QCD (SUSY-QCD)
one-loop contributions to the $t\rightarrow c\;h$ decays (\ref{tch}). In Sect. 4 we
address the SUSY-QCD contributions to $t\rightarrow c\;g$ in our framework
and compare with $t\rightarrow c\;h$. Finally, in Sect. 5 {we further discuss
our results} and deliver our conclusions.

\section{Relevant interaction Lagrangians}

\label{sec:gen}

The MSSM interaction Lagrangian involving fermions and $SU(3)_{c}\times
SU(2)_{L}\times U(1)_{Y}$ gauge bosons is well-known, and will not be
spelled out here in any detail\cite{MSSM,GJHS}. We will just focus on the
relevant interaction pieces entering the main one-loop contributions to the
Higgs channels (\ref{tch}).

To better present the analytic results of this computation we shall define
first a notation that allows us to treat the three possible decays~(\ref{tch}%
) in an unified way. We introduce a vector array of neutral Higgs fields 
\begin{equation}
\Phi ^{0}=(H^{0},h^{0},A^{0})\,\,,  \label{eq:defIndexH0}
\end{equation}
and another one for the charged Higgs and Goldstone bosons 
\begin{equation}
\Phi ^{+}=(H^{+},G^{+})\,\,.  \label{eq:defIndexHplus}
\end{equation}
Then the interaction Lagrangian of quarks with neutral and charged Higgs
bosons in the MSSM can be written 
\begin{eqnarray}
\mathcal{L}_{\Phi qq} &=&-\frac{g\,{m_u}}{2{M_W}%
{\sin\beta}}\sum_{r=1,3}\Phi _{r}^{0}\,\overline{u}%
\,(K_{r}^{0u}P_{L}+(K_{r}^{0u})^{\ast }P_{R})\,u\,  \nonumber \\
&&\,-\frac{g\,{m_d}}{2{M_W}{\cos\beta}}%
\sum_{r=1,3}\Phi _{r}^{0}\,\bar{d}\,(K_{r}^{0d}P_{L}+(K_{r}^{0d})^{\ast
}P_{R})\,d  \nonumber \\
&&+\frac{g}{\sqrt{2}{M_W}}V_{ud}\sum_{r=1,2}\left[ \Phi _{r}^{-}\,%
\bar{d}\,(K_{rL}^{+ud}P_{L}+K_{rR}^{+ud}P_{R})\,u+\mbox{ h.c.}\right]\,,
\label{Phiud}
\end{eqnarray}
with $P_{L,R}=$ $(1/2)\;(1\mp \gamma _{5})$ the chirality projectors. We
have defined the following sets of $K$-matrices: for the neutral Higgs
sector 
\begin{equation}
K_{r}^{0u}=\pmatrix{\sa \cr \ca \cr i\cbt}\,\,,\ \ K_{r}^{0d}=\pmatrix{\ca
\cr -\sa \cr i\sbt}\,,  \label{eq:Kdefinitionup}
\end{equation}
and for the charged Higgs sector 
\begin{equation}
\,\,K_{rL}^{+ud}={m_d}\pmatrix{\tb \cr -1}\,\,,\,\,\ \
K_{rR}^{+ud}={m_u}\pmatrix{\ctb \cr 1}\,\,.
\label{eq:Kdefinitiondown}
\end{equation}
In the above formula $V_{ud}$ is the CKM matrix element (assumed to be the
same as for the $W^{\pm }$ boson interactions with ordinary quarks).

The relevant Yukawa couplings involving charginos, quarks and squarks are
contained in 
\begin{equation}
\mathcal{L}_{u\,\sd\,\cplus}=-g\,V_{ud}\,\sd_{a}^{\ast }\,\bar{\psi}%
_{i}^{+}\left( \Apaidu\,\pl+\Amaidu\,\pr\right) \,u+\mbox{ h.c.}\,\,,
\label{eq:Lsqsqc}
\end{equation}
with $u$ (${\tilde d}$) up-type quarks (down-type squarks) of any
generation, the coupling matrices being\,\cite{GS1} 
\begin{equation}
\Apaidu=R_{1a}^{(d)\ast }V_{i1}^{\ast }-\lambda _{d}\,R_{2a}^{(d)\ast
}V_{i2}^{\ast }\,\,,\,\,\Amaidu=-R_{1a}^{(d)\ast }\lambda _{u}U_{i2}\,\,,
\end{equation}
where 
\begin{equation}
\lambda _{u}\equiv {\frac{h_{u}}{g}=\frac{m_{u}}{\sqrt{2}\,M_{W}\,\sin {%
\beta }}}\;\;\;\;\;,\;\;\;\;\;\lambda _{b}\equiv {\frac{h_{d}}{g}}={\frac{%
m_{d}}{\sqrt{2}\,M_{W}\,\cos {\beta }}}\,\,,  \label{eq:Yukawas}
\end{equation}
are the up-like and down-like Yukawa couplings normalized with respect to
the $SU(2)_{L}$ gauge coupling. Of fundamental importance for the SUSY-EW
enhancement of our FCNC decays through these Yukawa couplings is the value
of the parameter $\tan \beta =v_{2}/v_{1}$\cite{Hunter}. In the equations
above {$R^{(q)}$ }is the $2\times 2$ matrix that diagonalizes the squark mass
squared matrix in chiral space through 
\begin{equation}
\sq_{a}^{\prime }=\sum_{b}R_{ab}^{(q)}\sq_{b}^{{}}\,\,,  \label{R2}
\end{equation}
$\sq_{a}^{\prime }$ being the weak-eigenstates and $\sq_{a}$ the
mass-eigenstates.

Finally ({regarding the EW part}) we quote the interaction Lagrangian
for triplet Higgs vertices in the MSSM\footnote{%
Note that the elements $B_{rs3}$ are complex and $B_{ss3}=0$.} 
\begin{equation}
\mathcal{L}_{\Phi \Phi \Phi }=-g\sum_{r,s,t}B_{rst}\,\Phi _{r}^{+}\,\Phi
_{s}^{-}\,\Phi _{t}^{0}\,\,,  \label{eq:tripleHiggsGeneric}
\end{equation}
and the chargino couplings to neutral Higgs bosons 
\begin{equation}
\mathcal{L}_{{\chi^+}{\chi^+}\Phi }=-g\sum_{r,i,j}\Phi
_{r}^{0}\bar{\chi}_{i}^{+}(W_{ijL}^{r}\,P_{L}+W_{ijR}^{r}\,P_{R})%
{\chi^+_j}\,\,.  \label{eq:charcharHiggsgeneric}
\end{equation}
In both cases we have encapsulated the remaining notation in two $3\times 3$
matrices $B_{rst}$ and $W_{ij}^{r}$ whose explicit form can be identified
from Ref.~\cite{Hunter}. These matrices just give the corresponding Feynman
rules (divided my $-ig$).

On the other hand, the necessary SUSY-QCD interactions (in the mass-eigenstate basis) for our FCNC decays
are contained in the Lagrangian 
\begin{eqnarray}
\mathcal{L}_{\mathrm{SUSY-QCD}} &=&-\frac{g_{s}}{\sqrt{2}}\,\bar{\psi}_{c}^{%
{\tilde{g}}}\left[ R_{5\alpha }^{\ast }\,P_{L}-R_{6\alpha }^{\ast
}\,P_{R}\right] {\tilde q}_{\alpha ,i}^{\ast }\,\lambda
_{ij}^{c}\,t_{j}  \nonumber \\
&-&\frac{g_{s}}{\sqrt{2}}\,\bar{\psi}_{c}^{{\tilde{g}}}\left[
R_{3\alpha }^{\ast }\,P_{L}-R_{4\alpha }^{\ast }\,P_{R}\right] %
{\tilde q}_{\alpha ,i}^{\ast }\,\lambda _{ij}^{c}\,c_{j} 
\nonumber \\
&-&\frac{g_{s}}{\sqrt{2}}\,\bar{\psi}_{c}^{{\tilde{g}}}\left[
R_{1\alpha }^{\ast }\,P_{L}-R_{2\alpha }^{\ast }\,P_{R}\right] %
{\tilde q}_{\alpha ,i}^{\ast }\,\lambda _{ij}^{c}\,u_{j}+\mbox{ h.c.}\,\,,
\label{SUSYQCD}
\end{eqnarray}
where $\psi _{c}^{{\tilde{g}}}$ stands for the gluino spinor and $%
\lambda ^{c}$ are the $SU(3)_{c}$ Gell-Mann matrices. Moreover, the $6\times
6$ rotation matrices $R^{(q)}$ generalize those in eq.\,(\ref{R2}) and are
needed to diagonalize the squark mass matrices in (flavor)$\times $(chiral)
space as follows\footnote{%
First latin indices ($a,b,\ldots $) refer to different squark states of the
same flavor and take values $1,2$, with $\msqo<\msqt$ in the
mass-eigenstate basis and $\sq_{\{1,2\}}^{\prime }=\sq_{\{L,R\}}^{\prime }$
in the weak-eigenstate basis. Latin indices $i,j,\ldots =1,2$ refer to
charginos. $r,s,\ldots $ refer to Higgs and Goldstone particles. Greek
indices $\alpha ,\beta ,\ldots $ refer to squarks in the $6\times 6$ (flavor)%
$\times $(chiral) space and take values $1,2,\ldots ,6$, $\msqo<\msqt<\ldots <%
\msqs$.}: 
\begin{eqnarray}
\sq_{\alpha }^{\prime } &=&\sum_{\beta }R_{\alpha \beta }^{(q)}\sq_{\beta
}^{{}}\,\,,  \nonumber  \label{eq:definicioR6gen} \\
R^{(q)\dagger }\mathcal{M}_{\sq}^{2}R &=&\mathcal{M}_{\sq D}^{2}=\mathrm{diag%
}\{m_{\sq_{1}}^{2},\ldots ,m_{\sq_{6}}^{2}\}\,\,\,\,(q\equiv u,\,d)\,\,,
\end{eqnarray}
where $\mathcal{M}_{(\su,\sd)}^{2}$ is the $6\times 6$ square mass matrix
for squarks in the EW basis ($\sq_{\alpha }^{\prime }$), with indices $%
\alpha =1,2,3,\ldots ,6\equiv \su_{L},\su_{R},\tilde{c}_{L},\ldots ,\stopp%
_{R}$ for up-type squarks, and a similar assignment for down-type squarks.
However, in this study only up-type squarks are involved, so that we
understand that the above diagonalizing matrices refer to them.
The intergenerational mixing terms leading
to gluino-mediated FCNC couplings
lie in the off-diagonal entries of the mass matrices. However, in order to
prevent the number of parameters from being too large, we have allowed
(symmetric) mixing mass terms only for the left-handed (LH) squarks. This
simplification is often used in the MSSM and it is justified by
Renormalization Group (RG) analysis\,\cite{duncan}. Following this
practice, we introduce intergenerational (that is to say, flavor mixing)
coefficients $\delta _{ij}$
in the LL block of the mass matrix (namely the one involving only LH fields
of any flavor) as follows: 
\begin{equation}
(M_{LL}^{2})_{ij}=m_{ij}^{2}\equiv \delta _{ij}\,m_{i}\,m_{j}\ \ (i\neq
j)\,\,,  \label{eq:defdelta}
\end{equation}
where $m_{i}$ is the mass of the left-handed $i$th squark, and $m_{ij}^{2}$
is the mixing mass matrix element between generations $i$ and $j$.
Therefore, if the coefficients $\delta _{ij}$ are non-vanishing, for some $%
i\neq j$, then the structure of the diagonalizing matrices $R^{(q)}$ defined
above must necessarily lead to gluino-mediated tree-level FCNC between
quarks and squarks in the SUSY-QCD Lagrangian (\ref{SUSYQCD}). This scenario
can be generalized if we further introduce FCNC interactions on the
right-handed (RH) block of the mass matrix\cite{gabbiani,Pokorski} (see Sects.
4-5).

We note that the induced FCNC couplings in the SUSY-QCD Lagrangian (\ref{SUSYQCD})
ultimately stem from the fact that the squark mass matrix in general need
not to diagonalize with the same matrices as the quark mass matrix, i.e. the
so-called misalignment of quark and squark mass matrices.

\section{SUSY contributions to $t\rightarrow c\;h$}

From the previous interaction Lagrangians, the computation of FCNC processes
at one loop in a renormalizable theory is straightforward. In fact,
renormalization of parameters and Green functions is not needed, unlike the
usual flavor-conserving processes; one just computes the different diagrams
that contribute to the process and the final result obtained after adding up
all the amplitudes must be finite since no lowest order interaction could
absorb the left over infinities.

For every decay process (\ref{tch}) it is convenient to define an
``effective'' interaction vertex 
\begin{equation}
-i\,T\equiv -i\,g\;\,\overline{u}_{c}(p)\,\left(
F_{L}\,P_{L}+F_{R}\,P_{R}\right) \,u_{t}(k)\,\,,  \label{eq:effvertex}
\end{equation}
where $F_{L}$ and $F_{R}$ are form factors associated to each chirality
projector. They follow from explicit calculation of vertices and mixed
self-energies. 

The 
one-loop graphs for the decays under study are depicted in
Figs.\,\ref{diag:fcncEWvertex}, \ref{diag:fcncEWmix}
and~\ref{diag:fcncQCDvertex}. 
On the one hand the contribution from the vertex diagrams (Figs.\,\ref
{diag:fcncEWvertex} and~\ref{diag:fcncQCDvertex}a) 
is obtained by direct calculation. On the other hand for the (mixed)
self-energy diagrams (Cf. Figs.\,\ref{diag:fcncEWmix} and~\ref
{diag:fcncQCDvertex}b) 
it is convenient to define (in analogy with the 
charged-current case\,\cite{GJHS}) the following structure: 
\begin{equation}
\Sigma _{tc}(k)\equiv \rlap/k\,\Sigma _{L}(k^{2})\,P_{L}+\rlap/k\,\Sigma
_{R}(k^{2})\,P_{R}+m_{t}\,\left( \Sigma _{Ls}(k^{2})\,P_{L}+\Sigma
_{Rs}(k^{2})\,P_{R}\right) \,\,.  \label{eq:automixta}
\end{equation}
Here the $m_{t}$ factor multiplying the scalar part is arbitrary and it is
inserted there only to preserve the same dimensionality for the different $%
\Sigma _{i}$. From the Lagrangian interactions introduced in
Sect.\,\ref{sec:intro} we find that the effects of the mixed
self-energy diagrams to the amplitude of $t\rightarrow c\,\,\Phi _{r}^{0}$
take on the following general form in terms of the various $\Sigma _{i}$: 
\begin{eqnarray}
-i\,T_{S_{c}}^{r} &=&\frac{-i\,g\,{m_t}}{2{M_W}%
{\sin\beta}}\frac{1}{{m_c^2}-{m_t^2}}\bar{u}%
_{c}(p)\left\{ \phantom{ \pr\,(K^{0u}_{r})^*\,\left[ L\leftrightarrow R
\right] }\right.  \nonumber \\
&&P_{L}\,K_{r}^{0t}\,\left[ {m_c^2}\Sigma _{R}({m_c^2}%
)+{m_c}\,{m_t}\,\left( \Sigma _{Rs}({m_c^2}%
)+\Sigma _{L}({m_c^2})\right) +{m_t^2}\,\Sigma _{Ls}(%
{m_c^2})\right]  \nonumber \\
&&+\left. P_{R}\,(K_{r}^{0t})^{\ast }\,\left[ L\leftrightarrow R\right]
\right\} u_{t}(k)\,\,,  \nonumber \\
-i\,T_{S_{t}}^{r} &=&\frac{-i\,g\,{m_c}}{2{M_W}%
{\sin\beta}}\frac{{m_t}}{{m_t^2}-%
{m_c^2}}\bar{u}_{c}(p)\left\{ P_{L}\,K_{r}^{0c}\,\left[ %
{m_t}\left( \Sigma _{L}({m_t^2})+\Sigma _{Rs}(%
{m_t^2})\right) \right. \right.  \nonumber \\
&&\left. \left. +{m_c}\,\left( \Sigma _{R}({m_t^2}%
)+\Sigma _{Ls}({m_t^2})\right) \right] +P_{R}\,(K_{r}^{0c})^{\ast
}\,\left[ L\leftrightarrow R\right] \right\} u_{t}(k)\,\,,  \label{eq:selfs}
\end{eqnarray}
{where we have called }$T_{S_{i}}^{r}$ {the amplitude corresponding to
attaching the external }$r$th {Higgs particle line to the }$i$th {quark leg (%
}$r=1,2,3;\,i=c,t${)}. From~(\ref{eq:selfs}) 
the corresponding contribution to the form factors in eq.\,(\ref{eq:effvertex}%
) is transparent.

\subsection{SUSY-EW effects}

We start by reporting on the SUSY-EW effects, which by definition include
the sparticle contributions plus the Higgs-Goldstone boson diagrams 
computed in the Feynman gauge (Cf. Figs.\,\ref{diag:fcncEWvertex}
and~\ref{diag:fcncEWmix}). For the electroweak part we work in the so called
Super-CKM basis, that is, we take the simplification that the squark mass
matrix diagonalizes as the quark mass matrix, so that FCNC processes appear
at one-loop only through the charged current interactions (from charged Higgs and
charginos) and with the same mixing matrix elements as in the Standard Model
(the CKM matrix). It means that for the electroweak effects we do not take
into account the possible mismatch between the matrices diagonalizing the
squark sector and those diagonalizing the quark sector as this would only
result in a subleading additional correction.

Let us consider the computation of the contributions from 
diagrams in Figs.\,\ref{diag:fcncEWvertex} and~\ref{diag:fcncEWmix}. The graphs
in which FCNC are mediated by   
gauge bosons ($W^{+}$) have not been included since they are subdominant as
compared to the leading effects from the Yukawa couplings of the top and (at
large ${\tan \beta }$) bottom quarks. Still, the leading terms from
longitudinal $W^{+}$ are taken into account through the inclusion of
Goldstone bosons in the Feynman gauge.

The simplest SUSY-EW contributions are the self-energy diagrams
(Fig.\,\ref{diag:fcncEWmix}). They are obtained after substituting
the following expressions into the generic eq.\,(\ref{eq:selfs}): 
\begin{eqnarray}
\left. \Sigma _{R}(k^{2})\right| _{(a)} &=&i\,g^{2}\,V_{td}\,V_{cd}\,\Amaidc%
\,\Amaidt\,B_{1}(k,\mi,\msdal)  \,\,,\nonumber \\
\left. \Sigma _{L}(k^{2})\right| _{(a)} &=&i\,g^{2}\,V_{td}\,V_{cd}\,\Apaidc%
\,\Apaidt\,B_{1}(k,\mi,\msdal)  \,\,,\nonumber \\
\mt\,\left. \Sigma _{Rs}(k^{2})\right| _{(a)} &=&i\,g^{2}\,\mi%
\,V_{td}\,V_{cd}\,\Apaidc\,\Amaidt\,B_{0}(k,\mi,\msdal) \,\,, \nonumber \\
\mt\,\left. \Sigma _{Ls}(k^{2})\right| _{(a)} &=&i\,g^{2}\,\mi%
\,V_{td}\,V_{cd}\,\Amaidc\,\Apaidt\,B_{0}(k,\mi,\msdal)  \,\,,\nonumber \\
\left. \Sigma _{R}(k^{2})\right| _{(b)} &=&\frac{i\,g^{2}\,\mc\,\mt}{2\,\mws}%
V_{td}\,V_{cd}\left[ \ctbs\,\left( B_{0}+B_{1}\right) (k,\mHp,\mda)\right. 
\nonumber \\
&&\left. \phantom{\ctbs}+\left( B_{0}+B_{1}\right) (k,\mw,\mda)\right] 
\,\,,\nonumber \\
\left. \Sigma _{L}(k^{2})\right| _{(b)} &=&\frac{i\,g^{2}\,\mdas}{2\,\mws}%
V_{td}\,V_{cd}\left[ \tbs\,\left( B_{0}+B_{1}\right) (k,\mHp,\mda)\right. 
\nonumber \\
&&\left. \phantom{\tbs}+\left( B_{0}+B_{1}\right) (k,\mw,\mda)\right] 
\,\,,\nonumber \\
\mt\,\left. \Sigma _{Rs}(k^{2})\right| _{(b)} &=&\frac{i\,g^{2}\,\mt\,\mdas}{%
2\,\mws}V_{td}\,V_{cd}\left[ B_{0}(k,\mHp,\mda)-B_{0}(k,\mw,\mda)\right] 
\,\,,\nonumber \\
\mt\,\left. \Sigma _{Ls}(k^{2})\right| _{(b)} &=&\frac{i\,g^{2}\,\mc\,\mdas}{%
2\,\mws}V_{td}\,V_{cd}\left[ B_{0}(k,\mHp,\mda)-B_{0}(k,\mw,\mda)\right]
\,\,,
\end{eqnarray}
where the two-point functions $B_{i}$ are as in Ref.\cite{GJHS}. Of course
in the above expressions we understand a sum over the free indices $a,i=1,2$
and over the three {down-quark }generations ($d=d,s,b$).

The contribution to the form factors $F_{L}$ and $F_{R}$ in eq.\,(\ref
{eq:effvertex}) from SUSY-EW vertex diagrams is much more cumbersome.
Diagrams (a) and (d) of Fig.\,\ref{diag:fcncEWvertex} give a
generic contribution of the form 
\begin{eqnarray}
F_{L} &=&N_{A}\left[ ({C_{12}}-{C_{11}})%
{m_t}\,A_{R}^{(1)}\,A_{R}^{(2)}-{C_{12}}\,%
{m_c}\,A_{L}^{(1)}\,A_{L}^{(2)}+{C_0}%
\,m_{A}\,\,A_{R}^{(1)}\,A_{L}^{(2)}\right]  \,\,,\nonumber \\
F_{R} &=&F_{L}\,(A_{L}^{(\ast )}\leftrightarrow A_{R}^{(\ast )})\,\,,
\label{FLR1}
\end{eqnarray}
whereas diagrams\,(b) and (c) have the general structure 
\begin{eqnarray}
F_{L} &=&N_{D}\,\left[ C_{0}\,(D_{L}^{(1)}\,D_{L}^{(2)}\,D_{R}^{(3)}\,%
{m_c}\,{m_t}+D_{L}^{(1)}\,D_{L}^{(2)}\,D_{L}^{(3)}\,%
{m_c}\,m_{D1}\right.  \nonumber \\
&&\ \ \ \ +D_{R}^{(1)}\,D_{L}^{(2)}\,D_{R}^{(3)}\,{m_t}%
\,m_{D2}+D_{R}^{(1)}\,D_{L}^{(2)}\,D_{L}^{(3)}\,m_{D1}\,m_{D2})  \nonumber \\
&&+C_{12}\,{m_c}\,(D_{R}^{(1)}\,D_{R}^{(2)}\,D_{L}^{(3)}\,%
{m_c}+D_{L}^{(1)}\,D_{L}^{(2)}\,D_{R}^{(3)}\,{m_t} 
\nonumber \\
&&\ \ \
+D_{L}^{(1)}\,D_{L}^{(2)}\,D_{L}^{(3)}\,m_{D1}+D_{L}^{(1)}\,D_{R}^{(2)}%
\,D_{L}^{(3)}\,m_{D2})  \nonumber \\
&&+(C_{11}-C_{12})\,{m_t}\,(D_{L}^{(1)}\,D_{L}^{(2)}\,D_{R}^{(3)}%
\,{m_c}+D_{R}^{(1)}\,D_{R}^{(2)}\,D_{L}^{(3)}\,{m_t} 
\nonumber \\
&&\left. \ \ \
+D_{R}^{(1)}\,D_{R}^{(2)}\,D_{R}^{(3)}\,m_{D1}+D_{R}^{(1)}\,D_{L}^{(2)}%
\,D_{R}^{(3)}\,m_{D2})+{\tilde{C}_0}\,D_{R}^{(1)}\,D_{R}^{(2)}%
\,D_{L}^{(3)}\right]\,\,,  \nonumber \\
F_{R} &=&F_{L}(D_{L}^{(\ast )}\leftrightarrow D_{R}^{(\ast )})\,\,,
\label{FLR2}
\end{eqnarray}
where the three-point functions $C_{i}$ and $C_{ij}$ are as in Ref.\cite
{CGGJS,GJHS}.

Specifically, each vertex diagram of Fig.\,\ref{diag:fcncEWvertex}
contributes the following to the process $t\rightarrow c\,\,\Phi _{r}^{0}$:

\begin{itemize}
\item  Diagram\,(a); make the following substitutions in eq.\,(\ref{FLR1}%
): 
\begin{eqnarray*}
&&A_{L}^{(1)}=\Apbidc\,\,,\,\,A_{R}^{(1)}=\Ambidc\,\,,\,\,A_{L}^{(2)}=\Apaidt%
\,\,,\,\,A_{R}^{(2)}=\Amaidt \,\,,\\
&&m_{A}=\mi\,\,,\,\,N_{A}=i\,g^{2}V_{td}V_{cd}R_{ea}^{(d)}\,(R_{fb}^{(d)})^{%
\ast }\,G_{fe}^{r}\,\,,
\end{eqnarray*}
where $G_{fe}^{r}$ is the well-known Feynman rule\,\cite{Hunter}
for the vertex $\Phi _{r}^{0}\sd_{f}^{\prime }\,\sd_{e}^{\prime \ast }$
divided by $-ig$, in the electroweak-eigenstate basis\footnote{%
We recall that our sign convention\,\cite{CGGJS,GJHS,GS1} for the $\mu $
parameter is opposite to that of\,\cite{Hunter}.}. In this
diagram the various three-point functions on~(\ref{FLR1}) must be evaluated
with arguments 
\[
C_{\ast }=C_{\ast }(k,-p^{\prime },\mi,\msdal,\msdbe)\,\,. 
\]
The convention for the momenta can be seen in Fig.\,\ref
{diag:fcncEWvertex}a.

\item  Diagram\,(b); make the following substitutions in eq.\,(\ref
{FLR2}): 
\begin{eqnarray*}
&&D_{L}^{(1)}=\Apajdc\,\,,\,\,D_{R}^{(1)}=\Amajdc\,\,,\,%
\,D_{L}^{(2)}=W_{ijL}^{r}\,\,,\,\,D_{R}^{(2)}=W_{ijR}^{r} \,\,,\\
&&D_{L}^{(3)}=\Apaidt\,\,,\,\,D_{R}^{(3)}=\Amaidt \,\,,\\
&&m_{D1}=\mi\,\,,\,\,m_{D2}=\mj\,\,,\,\,N_{D}=i\,g^{2}\,V_{td}\,V_{cd} \,\,,\\
&&C_{\ast }=C_{\ast }(k,-p^{\prime },\msdal,\mi,\mj)\,,
\end{eqnarray*}

\item  Diagram\,(c); substitute in eq.\,(\ref{FLR2}): 
\begin{eqnarray*}
&&D_{L}^{(1)}=\kpcl\,\,,\,\,D_{R}^{(1)}=\kpcr\,\,,\,\,D_{L}^{(2)}=\kz%
\,\,,\,\,D_{R}^{(2)}=(\kz)^{\ast } \,\,,\\
&&D_{L}^{(3)}=\kptr\,\,,\,\,D_{R}^{(3)}=\kptr \,\,,\\
&&m_{D1}=m_{D2}=\mb\,\,,\,\,N_{D}=i\frac{g^{2}\,\md}{4\mw^{3}\,\cbt}%
V_{td}V_{cd} \,\,,\\
&&C_{\ast }=C_{\ast }(k,-p^{\prime },m_{\Phi _{s}^{+}},\md,\md)\,\,,
\end{eqnarray*}

\item  Diagram\,(d); substitute in eq.\,(\ref{FLR1}): 
\begin{eqnarray*}
&&A_{L}^{(1)}=\kpclj\,\,,\,\,A_{R}^{(1)}=\kpcrj\,\,,\,\,A_{L}^{(2)}=\kptl%
\,\,,\,\,A_{R}^{(2)}=\kptr \,\,,\\
&&m_{A}=\md\,\,,\,\,N_{A}=i\frac{g^{2}}{2\,\mws}\,B_{svr}\,V_{td}\,V_{cd} \,\,,\\
&&C_{\ast }=C_{\ast }(k,-p^{\prime },\md,m_{\Phi _{s}^{+}},m_{\Phi
_{v}^{+}})\,\,.
\end{eqnarray*}
\end{itemize}

As can be noted from the above expressions, the form factors induced by
Higgs mediated diagrams -- see Figs.\,\ref{diag:fcncEWvertex}c
and~d -- have the property $F_{L}=F_{R}$ for $H^{0}$ and $h^{0}$, and $%
F_{L}=-F_{R}$ for $A^{0}$.

We have performed the usual checks of the computation, in particular we find
that the form factors $F_{L}$ and $F_{R}$ are free of divergences before
adding up the three quark generations, both analytically and numerically in
the implementation of the code.

\subsection{SUSY-QCD effects}

Using the Lagrangians (\ref{Phiud}) and (\ref{SUSYQCD}) one can find the
SUSY-QCD one-loop contributions to the processes under study. They are much
more simple than the electroweak ones. The Feynman diagrams are depicted in
Figs.\,\ref{diag:fcncQCDvertex}a and~b. The one-loop mixed self-energy in
Fig.\,\ref{diag:fcncQCDvertex}b is determined from 
\begin{eqnarray}
\Sigma _{L}(k^{2}) &=&-i\,2\,\pi \,\alpha_{s}\,C_{F}\,R_{3\alpha
}\,R_{5\alpha }^{\ast }\,B_{1}(-k,{m_{\tilde{g}}},%
{m_{\tilde u_\alpha}}) \,\,, \nonumber  \label{eq:fcncQCDmixing} \\
\Sigma _{R}(k^{2}) &=&-i\,2\,\pi \,\alpha _{s}\,C_{F}\,R_{4\alpha
}\,R_{6\alpha }^{\ast }\,B_{1}(-k,{m_{\tilde{g}}},%
{m_{\tilde u_\alpha}}) \,\,, \nonumber \\
{m_t}\,\Sigma _{Ls}(k^{2}) &=&-i\,2\,\pi \,\alpha _{s}\,C_{F}\,%
{m_{\tilde{g}}}\,R_{4\alpha }\,R_{5\alpha }^{\ast }\,B_{0}(-k,%
{m_{\tilde{g}}},{m_{\tilde u_\alpha}}) \,\,, \nonumber \\
{m_t}\,\Sigma _{Rs}(k^{2}) &=&-i\,2\,\pi \,\alpha _{s}\,C_{F}\,%
{m_{\tilde{g}}}\,R_{3\alpha }\,R_{6\alpha }^{\ast }\,B_{0}(-k,%
{m_{\tilde{g}}},{m_{\tilde u_\alpha}})\,\,,
\end{eqnarray}
where $C_{F}=(N_{c}^{2}-1)/2N_{c}=4/3$ is the quadratic Casimir invariant of
the fundamental representation of $SU(3)_{c}$.

Finally, the SUSY-QCD vertex contributions to the form factors~(\ref
{eq:effvertex}) follow from Fig.\,\ref{diag:fcncQCDvertex}a, and
read 
\begin{eqnarray}
F_{L} &=&N\,\left[ {m_t}\,R_{4\beta }\,R_{6\alpha }^{\ast }\,(%
{C_{11}}-{C_{12}})+{m_c}\,R_{3\beta
}\,R_{5\alpha }^{\ast }\,{C_{12}}+{m_{\tilde{g}}}%
\,R_{4\beta }\,R_{5\alpha }^{\ast }{C_0}\right]  \,\,,\nonumber \\
F_{R} &=&F_{L}\left( 3\leftrightarrow 4\,\,,\,\,5\leftrightarrow 6\right)
\,\,,\label{FLRQCD}\nonumber \\
N &=&i\,8\,\pi \,\alpha _{s}\,C_{F}\,R_{\gamma \beta }^{\ast }\,G_{\gamma
\delta }^{r}\,R_{\delta \alpha }  \,\,,\nonumber \\
C_{\ast } &=&C_{\ast }(-k,p^{\prime },{m_{\tilde{g}}},%
{m_{\tilde u_\alpha}},{m_{\tilde u_\beta}})\,\,,
\end{eqnarray}
where $G_{\gamma \delta }^{r}$ is the well-known Feynman rule (divided by $%
-ig$) \cite{Hunter} for the vertex $\Phi _{r}^{0}{\tilde u}%
_{\gamma }^{\prime }{\tilde u}_{\delta }^{\prime \ast }$ in the
electroweak-eigenstate basis. From these expressions it should be clear that
if the up-type squarks would be degenerate, then by the unitarity of the $R$%
-matrices the gluino-mediated FCNC effects would vanish (GIM mechanism). We
have used this analytical property as an additional check of our numerical
code.

\section{Numerical Analysis of $t\rightarrow c\;h$}

\label{sec:numeric} \label{sec:ew}

After squaring the matrix element~(\ref{eq:effvertex}), and multiplying by
the phase space factor, one obtains the decay width of $t\rightarrow c\;h$, 
\begin{eqnarray}
\Gamma ({t\rightarrow c\,h}) &=&\frac{g^{2}}{32\,\pi \,%
{m_t}^{3}}\lambda ^{1/2}({m_t^2},m_{h}^{2},%
{m_c^2})  \nonumber \\
&\times &\left[ ({m_t^2}+{m_c^2}-m_{h}^{2}%
)(|F_{L}|^{2}+|F_{R}|^{2})+2\,{m_t}\,{m_c}%
\,(F_{L}\,F_{R}^{\ast }+F_{L}^{\ast }\,F_{R})\right] \,\,,
\label{eq:decaywidth}
\end{eqnarray}
with $\lambda(x^{2},y^{2},z^{2})=(x^{2}-(y+z)^{2})(x^{2}-(y-z)^{2})$ the usual
K{\"a}llen function. It is conventional to define the ratio 
\begin{equation}
B({t\rightarrow c\,h})\equiv \frac{\Gamma (t\rightarrow c\,h)}{%
\Gamma (t\rightarrow b\,W^{+})}\,,  \label{eq:defbr}
\end{equation}
which will be the main object of our numerical study.

This ratio is not the total branching fraction $BR(t\rightarrow c\,h)$ of
the decay mode, as there are many other channels that should be added up to
the denominator of (\ref{eq:defbr}) in the MSSM, if kinematically allowed,
such as the two and three body decays of the top quark into SUSY particles,
and also the charged Higgs decay channel $t\rightarrow
H^{+}\,b$\,\cite{CGGJS,GS1}. For the mass spectrum used in our numerical
analysis the 
former decays are phase space closed, whereas the latter could have a
sizeable branching ratio. However, for better comparison with previous
analyses of FCNC top quark decays \,\cite{Mele}-\cite{Divitiis}, the fiducial
quantity (\ref{eq:defbr}) 
should suffice to assess
the experimental viability of the FCNC decays under consideration.

In the following we will analyze the numerical contributions to (\ref
{eq:defbr}) from the SUSY-EW and SUSY-QCD sectors, in a sparticle mass model
motivated by model building and RG analysis. However, we do not restrict
ourselves to the spectrum of specific SUSY-GUT models -- such as SUGRA or
constrained MSSM models{\cite{MSSM,SUGRA}. Furthermore, as 
announced in the beginning, all over our numerical analysis
we use {the full structure }of
the one-loop relations for the parameters in the Higgs sector of the
MSSM\,\cite{Dabelstein}.

{We start with the EW effects. Although we have generally scanned
the MSSM parameter space up to }$1\,TeV$ {level, the following
input set has been chosen where the individual parameters have to be fixed
at particular values to illustrate our results (as in Fig.\,\ref{fig:resultsew}):} 
\begin{equation}
\begin{array}{l}
{\tan \beta }=35\,\,,\,\,\mu =-500{\,GeV}\,\,,\,\,M=150{\,GeV}%
\,\,,\,\,M_{A^{0}}=100{\,GeV}\,\,, \\ 
m_{\tilde{t}_{1}}=150{\,GeV}\,\,,\,\,m_{\tilde{b}_{1}}=m_{\tilde{q}%
}=200{\,GeV}\,\,,\,\,A_{t}=A_{q}=300{\,GeV}\,\,,\,\,A_{b}=-300%
{\,GeV}\,\,.
\end{array}
\label{eq:inputew}
\end{equation}
We have taken the third generation quark masses as $\mt=175\,GeV$ and $\mb=5\,GeV$.
In the previous equation $m_{{\tilde{t}}_{1}},m_{{\tilde{b}}_{1}}$ are the lightest ${\tilde{t}}$
and ${\tilde{b}}$ mass, and all the masses are above present experimental
bounds\,\cite{UAB97}. However, we have chosen a SUSY mass spectrum around
$200{\,GeV}$, which is not too light, so the results will not be 
artificially optimized. We have also checked that all through our numerical
analysis other bounds on experimental parameters (such as $\delta \rho $)
are fulfilled.

We have found that the contributions to the form factors of~(\ref{eq:effvertex})
are of the same order for the chargino (Figs.\,\ref{diag:fcncEWvertex}a,b
and~\ref{diag:fcncEWmix}a) and Higgs 
particles\,\cite{JGI} (Figs.\,\ref{diag:fcncEWvertex}c,d
and~\ref{diag:fcncEWmix}b) --not included in previous analyses~\cite{tch}.
It turns out that they can be either of the same sign, or of opposite sign.
The chosen negative value for $A_{b}$ is to make the two contributions of
the same sign. In both cases {$F_{R}\gg F_{L}$.} This can be easily
understood by looking at the interaction vertices involving
(charged)Higgs-bottom-charm\,\cite{Hunter} and chargino-sbottom-charm, where the latter
can be tracked down
from the explicit structure of the Lagrangian~(\ref{eq:Lsqsqc}). In both of
them the contribution to the right-handed form factor will be enhanced by
the Yukawa coupling of the bottom quark, whereas the charm Yukawa coupling
contributes to the left-handed form factor. We have checked that the
inclusion of the first two generations of quarks and squarks only has an
effect of a few percent on the total result.

In Fig.\,\ref{fig:resultsew} we can see the evolution of the ratio~(%
\ref{eq:defbr}) with various parameters of the MSSM by taking into account
only the electroweak contributions. The growing of the width with $\tan
\beta $ (Cf. Fig.\,\ref{fig:resultsew}a) makes evident that the
bottom quark Yukawa coupling plays a central role in these contributions.
The evolution with the trilinear coupling $A_{b}$ {--}the main
parameter that appears in the interaction vertex $\tilde{b}_{L}\,\tilde{b}%
_{R}\,h$-- displayed in Fig.\,\ref{fig:resultsew}b shows that this
parameter can enhance the width by some orders of magnitude. On the other
hand the evolution with he higgsino mass parameter $\mu $ (Cf.
Fig.\,\ref{fig:resultsew}c) is {comparatively mild in the
region away from the origin }$\mu =0${. }The shaded region in
Fig.\,\ref{fig:resultsew}c, centered at the origin, is ruled out 
{by present LEP bounds on chargino masses\,\cite{UAB97}. As the
dependence of the rate (\ref{eq:defbr}) on the }$SU(2)_{L}$ {%
gaugino mass }$M$ {is essentially flat (not shown), these LEP bounds
effectively translate in our case into excluding} $\mu \lsim90\,GeV$.  The
various spikes in these figures reflect the points where the overall
numerical contribution from the form factors {(vertex plus self-energy)
  cancels out and even} changes sign. The actual point where this cancellation
occurs depends on the particular choice of the parameters.

The fact that in all these figures the ratio~(\ref{eq:defbr}) is smaller for
the heaviest scalar Higgs ($H^{0}$) is not due to the smaller phase space
factor, but to the smallness of the form factors. We can see in Fig.\,\ref
{fig:resultsew}d that in fact  $B(t\rightarrow c\,H^{0})$ grows with the $%
A^{0}$ (and thus the $H^{0}$) mass, until the phase space begins to close.

We conclude that a {typical value of the ratio~(\ref{eq:defbr}), at
large }$30\lsim \tb\lsim 50$ and for a SUSY spectrum around $200%
{\,GeV}$, reads roughly 
\begin{equation}
B^{\mathrm{SUSY-EW}}({t\rightarrow c\,h})\simeq 10^{-8}\,\,\,,
\label{eq:conew}
\end{equation}
provided $M_A<120-130\,GeV$.
This is larger than the previous reported ratios \cite{tch} by $2$ orders of
magnitude, specially in the $A^{0}$ channel, {and it is at least }$5$ {%
orders of magnitude larger than the SM rate } $B(t\rightarrow
c\,H_{SM})$\,{\cite{Mele,Erratum}. In fact, this feature can be explicitly
  checked in our 
framework in the limit }$M_{A^{0}}\rightarrow \infty $ {in which the CP-even
Higgs boson }$h^{0}$ {behaves like the SM Higgs boson }$H_{SM}$. {This is
already seen in part in Fig.\,\ref{fig:resultsew}d. By further
sending the SUSY masses to infinity we indeed recover a very poor FCNC rate
for }$h^{0}\sim H_{SM}$ {which goes down }$10^{-13}${\cite{Mele,Erratum}.}

Turning now to the analysis of the SUSY-QCD effects, it is clear that they
hinge to a great extent on the values of the flavor mixing
coefficients $\delta _{ij}$. The latter are constrained by low-energy data
on FCNC.  The bounds have been computed using some approximations, so they
must be taken as order of magnitude limits rather than as accurate numbers.
They read as follows \cite{gabbiani,Pokorski}: 
\begin{eqnarray}
|\delta _{12}| &<&.1\,\sqrt{m_{\tilde{u}}\,m_{\tilde{c}}}/{500{\,GeV}}
\,\,,\nonumber \\
|\delta _{13}| &<&.098\,\sqrt{m_{\tilde{u}}\,m_{\tilde{t}}}/{500{\,GeV%
}}\,\,,  \nonumber \\
|\delta _{23}| &<&8.2\,m_{\tilde{c}}\,m_{\tilde{t}}/{(500{\,GeV})^{2}}%
\,\,\,.  \label{eq:limdelta}
\end{eqnarray}
In using these bounds we make use of $SU(2)$ gauge invariance to transfer
the experimental information known from the down-quark sector (for example
from {$BR(b\rightarrow s\;\gamma )$}, where the bound on $\delta _{23}$ is
obtained) to the up-quark sector. It means that after soft SUSY breaking,
but before SSB, the LL blocks of the up-squark and down-squark mass matrices
must satisfy the following relation \cite{Pokorski} 
\begin{equation}
(M_{\widetilde{U}}^{2})_{LL}=K\;(M_{\widetilde{D}}^{2})_{LL}\;K^{\dagger }\,\,,
\end{equation}
where $K$ is the Cabibbo-Kobayashi-Maskawa matrix. Thus if $M_{\widetilde{D}%
}^{2}$ is parametrized as in eq.\,(\ref{eq:defdelta}), then $(M_{\widetilde{U}%
}^{2})_{LL}$ inherits a similar form with a new set of mixing coefficients $%
\delta _{ij}$which differ from the previous ones by factors of $\mathcal{O}%
(1)$.

We use the same input parameters as in the electroweak contributions~(\ref
{eq:inputew}), 
plus the specific parameters of the SUSY-QCD sector,  {namely the
gluino mass }$m_{\tilde{g}}$ {and the mixing coefficients }$\delta
_{ij}${, as follows}: 
\begin{eqnarray}
m_{\tilde{g}} &=&200\,{\,GeV\,,}  \nonumber \\
\delta &=&\left( 
\begin{array}{ccc}
0 & 0.03 & 0.03 \\ 
0.03 & 0 & 0.4 \\ 
0.03 & 0.4 & 0
\end{array}
\right) \,\,\,.  \label{eq:inputqcd}
\end{eqnarray}
As for the strong coupling constant we have used the value
$\alpha_s(\mts)=0.11$. A comment is in order for the present set of inputs: we
have introduced in~(\ref{eq:inputew}) the lightest stop mass as an input, and
this stop is 
mostly a ${\tilde{t}}_{R}$. However, in this new parametrization we
introduce this mass as the lightest ${\tilde{u}}_{\alpha }$ mass, which
again will be mostly a ${\tilde{t}}_{R}$.  {Notice furthermore that
the chosen entries in the mixing matrix (\ref{eq:inputqcd}) are moderate in
the sense that they do not saturate the permissible upper limits (\ref
{eq:limdelta}). Here we are using the additional constraint that the squark
masses (which are obviously affected by the values of the parameters }$%
\delta _{ij}$) {cannot be too light.  As mentioned above, although
we use a fixed ``typical'' choice of inputs, a systematic scanning has been
performed over the parameter space. In particular, the maximum values for
the rates (to be compared with the typical ones) have also been pinned down
(see later on).}

In analyzing the SUSY-QCD effects we find that again the largest contribution
comes from the right-handed form factor in eq.\,(%
\ref{eq:effvertex}), but this is only because {up to now} we have
chosen not to introduce mixing between right-handed squarks.
We have plotted the evolution of the ratio~(\ref{eq:defbr}) with some
parameters of the MSSM in Fig.\,\ref{fig:resultsqcd}. As it reads
off  eq.\,(\ref{eq:inputqcd}) the most relevant parameter in the SUSY-QCD
analysis is the mixing coefficient between the 2nd and 3rd generation
of LH squarks, which is the less restricted one of the three in eq.\,(\ref
{eq:limdelta}). In Fig.\,\ref{fig:resultsqcd}a it is shown that by
changing $\delta _{23}$ by 3 orders of magnitude the ratio~(\ref{eq:defbr})
increases by 6 orders of magnitude! -- {a fact that can be traced to the
quadratic dependence on the mixing coefficient}. {Worth noticing in
Fig.\,\ref{fig:resultsqcd}b is the $\mu $ parameter dependence of
the SUSY-QCD effects, which enters through the $\sq_{L}\sq_{R}h$ coupling.
The ratio (\ref{eq:defbr})} can be pushed up to values of $10^{-5}$ {%
irrespective of the sign of} $\mu $. Again the central region of $\mu $ is
excluded by present LEP bounds on the chargino masses.

The evolution with the gluino mass (Cf. Fig.\,\ref{fig:resultsqcd}%
c) is asymptotically quite stable, showing a slow decoupling. {Thus,
even for gluinos as heavy as }$500\,GeV$ {the rate for the top quark
decay into the lightest CP-even Higgs boson (}$t\rightarrow c\,h^{0}$)
{can stay above }$10^{-5}$. Finally in Fig.\,\ref{fig:resultsqcd}d we 
have plotted the evolution with the pseudoscalar Higgs 
mass. It is seen a behavior similar to the EW case (Cf.
Fig.\,\ref{fig:resultsew}d) although scaled up a factor $10^{2}-10^{3}$.
Obviously, the one-loop Higgs mass relations play an essential role here
(missed in \cite{tch})
in that the $h^0$ is bound to have a mass (below $\sim 135\,GeV$) which is higher
than in the tree-level case (below $M_Z$). Still the asymptotic behavior of 
$B(t\rightarrow c\,h^0)$ versus $M_{A^0}$
is sustained a long while around $10^{-6}$ for a SUSY spectrum of 
a few hundred GeV.

{Figs.\,\ref{fig:resultsqcdmasses}a and b display} the
change of~(\ref{eq:defbr}) with the lightest stop and sbottom masses within
the allowed region. It is clear that $\msto$ plays an important role, which
is due to the fact that by changing this parameter one is also changing the
value of the mixing angle between the LH and RH stops. On the other hand the
actual value of $\msbo$ is seen not to be that important {in the $%
A^{0}$ and $h^{0}$ channels}.

In Fig.\,{\ref{fig:resultsqcdmasses}c} we plot the ratio~(%
\ref{eq:defbr}) versus $\tb$. 
We see that the evolution of the SUSY-QCD contributions as a function of $%
\tb
$ is quite stable for the $h^{0}$ and $A^{0}$ channels, and so it does not matter
much, in the SUSY-QCD case, whether we are in the low or in the high $\tb$ region. The pronounced
downwards spike around $\tb\simeq 10$ for the $H^{0}$ channel is due to the
change of sign of the form factor.

For completeness we have explicitly probed the impact on the FCNC rates in
the presence of
intergenerational {mass-mixing} terms in the right-handed
sector of the model. In order to maintain the number of parameters under
control we have set $(\delta _{ij})_{RR}=(\delta _{ij})_{LL}$. We find that
the inclusion of the additional mixing coefficients does increase the FCNC
rate, but we do not plot the result since it just amounts to a total
contribution which is at most twice the old result with $(\delta
_{ij})_{RR}=0$. {And this feature holds for any set of input parameters~(\ref
{eq:inputew},\ref{eq:inputqcd}). Therefore, we conclude that the old
coefficients }$(\delta _{ij})_{LL}$ {alone} --{the only ones that are
naturally generated within RG-based models \cite{duncan} -- already bring
about the bulk of the FCNC rates (\ref{eq:defbr}).}

We wish to emphasize that although we have used a common subset of inputs --see
eq.\,(\ref{eq:inputew})-- to compute the SUSY-EW and the SUSY-QCD loops, 
the enhancement sources for the two types of contributions are entirely different.
Thus, whereas the SUSY-EW effects are much sensitive to extreme values of
the parameter $\tan \beta $ through the Yukawa couplings~(\ref{eq:Yukawas}),
the SUSY-QCD effects (which are the leading ones in our calculation) are not
particularly sensitive to $\tan \beta $, as confirmed in Fig.\,\ref
{fig:resultsqcdmasses}c. Instead, they are basically dependent on the
experimentally allowed values of the flavor mixing coefficients (\ref{eq:defdelta}%
), as it is plain in Fig.\,\ref{fig:resultsqcd}a.

From the previous numerical analysis we confirm that the preliminary MSSM
treatment of the FCNC decay of the top quark {into Higgs
  bosons\,\cite{tch} was fairly incomplete since important effects from Higgs
  particles in 
the loops were not included, and moreover the }$\tilde{q}_{L}\,\tilde{q}%
_{R}\,h$ {vertices were not taken into account. As a result the
potentially large contributions coming from the trilinear soft SUSY-breaking
terms }$A_{t,b}${, and from the higgsino mass parameter $\mu$ were
missed}. Moreover, the pattern of quantum effects in the three Higgs channels is
affected in an essential way by the one-loop Higgs mass relations of the MSSM.
From our rigorous computation we have been able to 
show that $B{(t\rightarrow c\,h)}$ in the MSSM can typically be of
order $10^{-8}$ for the electroweak contributions, and reach $10^{-5}$
for the QCD contributions. In some of the channels
this amounts to having rates that are one to two orders of magnitude
larger than previous estimates. For an assessment of the impact of these 
results on experimental searches, and a discussion of the maximum
attainable rates, see Sect. 6. 

\section{The decay $t\rightarrow c\;g$}

We have already mentioned that the FCNC decays of the top quark into gauge
bosons, $t\rightarrow c\;V$, may receive important contributions from SUSY
physics, and this has been confirmed by explicit calculations in the
literature \cite{Yang}{-\cite{Divitiis}}. {In particular}, the
situation with the decays $t\rightarrow c\,V$ into electroweak gauge bosons $%
V=\gamma ,Z$ is that, in spite of the SUSY enhancements, the final rates are
far insufficient to be seen, except for very especial circumstances
associated to the possibility of wave-function threshold effects \cite{Lopez}%
. These effects, however, are ``point-like coincidences'', {so to
speak}, which we deem to be very unlikely. Indeed, a similar situation
occurs e.g. when computing the SUSY corrections to the conventional decay of
the top quark, $t\rightarrow W^{+}b$ \,\cite{GJHS}. Also in this
case there are particular combinations of  sparticle masses which fall
within the (very) narrow range where wave-function threshold enhancements
occur\footnote{%
See Ref.\,\cite{GJHS} for details, and in particular
Figs.\,6, 7 and 8 of that reference.}. To be sure, this circumstance
should be considered exceptional and so the fairest conclusion ought rather
to be that the SUSY effects on $t\rightarrow W^{+}b$ are generally small,
and similarly that the SUSY rates for the FCNC decays $t\rightarrow
c\,(\gamma ,Z)$ are far below the experimental possibilities.

In the specific case of the gluon channel $t\rightarrow c\;g$ one could also
argue that it could be enhanced by threshold effects, only if the top quark,
gluino and stop masses turned out to satisfy the peculiar relation $%
m_{t}\simeq m_{\widetilde{t}}+m_{\widetilde{g}}$.  Nonetheless a relation
like this is not only contrived, but it is already ruled out by the current
bounds on gluino and stop masses~\cite{UAB97}: $m_{\widetilde{g}}>180\;GeV$, 
$m_{\widetilde{t_{1}}}>80\;GeV$. Fortunately, in contrast to the electroweak
gauge boson channels just mentioned, alternative important corrections to $%
t\rightarrow c\;g$ are possible from gluino-mediated FCNC loops. This
subject has recently been addressed in Ref.\,\cite{Divitiis}, but
due to discrepancies of this reference with previous calculations by the
authors of Refs.\cite{Yang,Koenig} we wish to reanalyze this decay in our
framework. In this way we {hope to further clarify the situation and
at the same time to use} $t\rightarrow c\;g$ as a fiducial observable with
which to better compare the results that we find for the $t\rightarrow c\;h$
channels within one and the same set of assumptions.

The diagrams contributing to $t\rightarrow c\;g$ are similar to those in
Figs.\,\ref{diag:fcncEWvertex}a,c and~\ref{diag:fcncQCDvertex}
after replacing the Higgs boson $h$ with the gluon $g$. {The corresponding
SM contribution involves the }$W$ {gauge boson and the bottom quark flowing
in the loop. } However, in the conditions of the foregoing study of $%
t\rightarrow c\;h$, we have seen that the set of electroweak diagrams in
Figs.\,\ref{diag:fcncEWvertex}-\ref{diag:fcncEWmix} furnish in
general a negligible contribution as compared to the gluino-mediated
contributions in Fig.\,\ref{diag:fcncQCDvertex} -- and we find that
this does not change for the decay $t\rightarrow c\;g$. Therefore, we only
report on the corresponding SUSY-QCD effects.

A crucial point in computing this type of effects in the case of $%
t\rightarrow c\,h$ was to realize that there are diagrams in the electroweak
eigenstate-basis with an helicity flip at the gluino line. As a result the
decay $t\rightarrow c\;h$ received contributions ``proportional'' to the
gluino mass --Cf. eq.\,(\ref{FLRQCD}). These terms cause the corresponding
decay rate to fall off very slowly with $m_{\widetilde{g}}$, and they even
produce a local maximum with respect to this parameter (recall
Fig.\,\ref{fig:resultsqcd}c). Similarly, it turns out that the
decay $t\rightarrow c\;g$ can also have this kind of enhancements, although
from a slightly different origin {(see below)}. This fact, with which
we agree with the recent calculation of Ref.\cite{Divitiis}, was missed in
Refs.\,\cite{Yang,Koenig} and it led them {to speculate on
the existence of additional gluino-mediated FCNC interactions in the RH
sector in order to reach higher rates.  Although these additional terms
are possible, in principle, they are unnatural in a RG-based framework; and
what is more, they are actually
unnecessary to potentially achieve the desired enhancements}.

The analytical calculations for $t\rightarrow c\;g$ are similar to those for 
$t\rightarrow c\;h$, and so we just report on the numerical analysis, which
we present in Figs.\,\ref{fig:tcgd23mg}-\ref{fig:tcgmstAt}. In all
these figures we exhibit the evolution of the ratio $B(t\rightarrow c\;g)$ -- 
{defined as in eq.\,(\ref{eq:defbr}) with }$g$ {replacing }$h$
-- for two different scenarios, namely, when only mixing between LH
squarks is allowed (solid lines), and when mixing mass terms are allowed
also in the RH squark sector (dashed lines). {If one sticks to the
more conservative FCNC gluino-mediated interactions in the pure LH sector,
we see that one may reach important enhancements which can bring }$%
B(t\rightarrow c\;g)$ {up to }$10^{-6}${-}$10^{-5}$ {%
for allowed sparticle masses\footnote{When comparing with Ref.\,\cite{Divitiis},
in the unconstrained MSSM case, we
point out that our rates are smaller only because we use an updated (larger) value of the
gluino mass and also a larger value for the lightest stop mass
-- see eq.\,(\ref{eq:inputew})-- which is the most sensitive one.
 }.}  However, {as in the }$t\rightarrow c\;h$
case, {we find } {that the inclusion of FCNC in the RH sector do not modify
the order of magnitude of the results; in fact the RH interactions only
enhance the previous result by at most a factor of }$2$.

{In the calculation, when including the RH terms, we have set }$%
(\delta _{ij})_{RR}=(\delta _{ij})_{LL}$ {and used the fact that the
bounds on }$(\delta _{ij})_{LL}$ {are interwoven with the bounds on
the squark masses \,\cite{Pokorski} as dictated by eq.\,(\ref
{eq:defdelta}), similarly as we did in our analysis of the }$t\rightarrow
c\;h$ {decays in Sect.\,3. }

In Fig.\,\ref{fig:tcgd23mg} we display the evolution of $%
B(t\rightarrow c\,g)$ as a function of the mixing parameter $\delta _{23}$
and the gluino mass $\mg$. 
We see that $B(t\rightarrow c\,g)$ is also proportional to
$(\delta_{23})^2$ as $B(t\to c\,h)$. The evolution with the gluino mass also exhibits
a local maximum, characteristic of the terms ``proportional'' to $\mg$ that
are triggered by the helicity flip in the gluino line. As we said before, {%
this flip occurs even if we only allow intergenerational mass-mixing terms
in the LL block of the mass matrix -- a feature which was completely
overlooked in Ref.\cite{Koenig}. }{The origin of the flip in the
electroweak-eigenstate basis is as follows}{. In the one-loop vertex diagram
there exists a FCNC gluino
interaction with the LH charm quark and the LH stop, where the latter stems
from a RH stop that has mutated (through a mixed propagator) into a LH
state. This situation is favorable because of large LR mixing in the stop
sector. Schematically, one can think in terms of the
following picture: from the gluon vertex in the loop there emerges a }$\stopp%
_{R}$ {that subsequently undergoes the series of transitions }$\stopp%
_{R}\rightarrow \stopp_{L}\rightarrow c_{L}$~\footnote{%
The difference with the $t\rightarrow c\,h$ case lies in the fact that the
first transition
$\stopp_{R}\rightarrow \stopp_{L}$ is already possible
at the Higgs vertex $\tilde{t}_{L}\,\tilde{t}_{R}\,h$ in the loop, but 
it cannot take place at
the gluon vertex.}. {As the structure of this loop enforces a mass
insertion in the gluino propagator, it leads to an enhancement
of }$B(t\rightarrow c\,g)$ {which} {nevertheless falls off with the gluino
mass much faster than in the }$B(t\rightarrow c\,h)$ {case (Cf.
Fig.\,\ref{fig:resultsqcd}c).}

The evolution with the electroweak parameters $\tb$ and $\mu $ is presented
in Fig.\,\ref{fig:tcgmutb}. In this process they only enter as
inputs in the squark mass matrix, but not in the couplings, so the
dependence of $B(t\rightarrow c\,g)$ on them is rather mild. In the presence
of mixing in both the LH and RH sectors, the physical squarks masses are
smaller than the corresponding ones with mixing only in the LH sector.
Eventually these particles can be lighter than present bounds, thus the
parameter space is further constrained. This is shown in Fig.\,\ref
{fig:tcgmutb} where the dashed lines (corresponding to $\delta _{LL}=\delta
_{RR})$ are cut off at points in which the present bounds on squarks masses
would be violated. Finally, we report in Fig.\,\ref{fig:tcgmstAt}
on the evolution with the lightest stop mass and the soft SUSY-breaking
trilinear coupling $A_{t}$. The shaded region in Fig.\,\ref
{fig:tcgmstAt}b is excluded as it incompatible with our choice of
parameters, and the
various lines end up where the large mixing would induce too light squark
masses. The rate is seen to be sensitive to both parameters $m_{\tilde{t}_1}$
and $A_{t}$, but the
actually permitted region for the latter is quite narrow.

{To summarize, upon comparing the decay }$t\rightarrow c\;g$ {with
  the relevant decays }$t\rightarrow c\;h$ {under study we find 
that the latter can be relatively very important, and most likely they are
the dominant FCNC top quark decays in the MSSM. As a matter of fact, whereas
the critical level }$10^{-5}$ {is never surpassed by the gluon
channel -- even with LH+RH effects and for sparticle masses within the
current limits--, the }$t\rightarrow c\;h^{0}$ {and }$t\rightarrow
c\;A^{0}$ {modes, instead, may well crossover the visible level
already with the more conventional LH contributions. The maximum values for
all these decays are discussed in the next section, where we also present
our conclusions.}

\section{Discussion and conclusions}

\label{sec:con}

We have studied the SUSY-EW and SUSY-QCD contributions to the {leading%
} FCNC top quark decays. We have mainly {concentrated on the top
quark FCNC decays into the Higgs bosons of the MSSM}, namely ${t\rightarrow
c\,h}$  ($h=h^{0},H^{0},A^{0}$), using a mass spectrum motivated, but not
fully restricted, by model building and Renormalization Group Equations. And
we have found that a full treatment of the SUSY-QCD contributions may
greatly enhance the FCNC width by some orders of magnitude {with
respect to the FCNC decay rates into gauge bosons }$t\rightarrow
c\,V\,(V\equiv \gamma ,Z,g)$. Due to the crucial role played by the SUSY-QCD
effects, we have reconsidered the FCNC decay into gluons, ${t\rightarrow c\,g%
}$, which is the most promising one among the FCNC top quark decays into
gauge bosons. And we have shown that under similar assumptions of
gluino-mediated FCNC couplings in the LH sector, it could be enhanced up to 
{near} the visible level, but below the rates of the Higgs channels.
The additional RH interactions, if present at all, could produce a further
increase of the rates, but it just amounts a factor of $2$ at most. The
remaining gauge boson channels give maximum rates in the MSSM that are far
below experimental possibilities, except in highly unlikely circumstances.

With a SUSY mass spectrum around $200{\,GeV}$, which is {above
the current absolute LEP bounds} \cite{UAB97}, the different contributions
to the Higgs channels are typically of the order 
\begin{eqnarray}
B^{\mathrm{SUSY-EW}}({t\rightarrow c\,h}) &\simeq &10^{-8}\,\,,  \nonumber
\label{eq:confinal} \\
B^{\mathrm{SUSY-QCD}}({t\rightarrow c\,h}) &\simeq &10^{-5}\,\,.
\end{eqnarray}
However, by stretching out a bit more the range of parameters one can reach (%
{for some of the decays}) 
\begin{eqnarray}
B^{\mathrm{SUSY-EW}}({t\rightarrow c\,h}) &\simeq &1\times 10^{-6} \,\,, \nonumber
\\
B^{\mathrm{SUSY-QCD}}({t\rightarrow c\,h}) &\simeq &5\times 10^{-4}.
\label{corner}
\end{eqnarray}

The difference of at least two orders of magnitude between the SUSY-EW and
SUSY-QCD contributions makes unnecessary to compute the interference terms
between the two sets of amplitudes, but if the limits on $\delta _{23}$
(eq.\,(\ref{eq:limdelta})) become eventually more restrictive then they
should be taken into account.

{We have obtained the maximum rates (\ref{corner}) from a general
search in the MSSM parameter space within the }$1\,TeV$ {mass
region. }In Fig.\,\ref{fig:maxims} we present the maximum values
that can be reached by $B(t\rightarrow c\,X)$ for each of the processes
presented. In Figs.\,\ref{fig:maxims}a and~b we show the maximized $%
B(t\rightarrow c\,h)$ as a function of the pseudoscalar Higgs boson mass by
taking into account only the SUSY-EW contributions and the SUSY-QCD
contributions respectively. We have performed a systematic scanning of the
parameter space of the MSSM, with the various masses constrained between the
present exclusion bounds and the $1\,TeV$ upper bound, and using the other
bounds from Sect.\,\ref{sec:numeric} -- in particular, see
eq.\,(\ref{eq:limdelta}). 
Perhaps the most noticeable result is that the decay into the
lightest  
MSSM Higgs boson ($t\rightarrow c\,h^{0}$) is the one that can be maximally
enhanced and reaching values of order 
$B(t\rightarrow c\,h^{0})\sim 10^{-4}$ that stay fairly stable
all over the parameter space. The reason for
this dominance is that the decay $t\rightarrow c\,h^{0}$ is the one which is
more sensitive to $A_{t}$, a parameter whose natural range reaches up to
about $1\,TeV$. For more moderate values of $A_{t}$, however, the dominant
Higgs decay mode can be $t\rightarrow c\,H^{0}$, but in this case the
corresponding rate (also that of $t\rightarrow c\,A^{0}$) undergoes a rapid
fall-off with $M_{A^{0}}$ (Cf. Fig.\,\ref{fig:maxims}b) as it was also the
case in the unmaximized situation in Fig.\,\ref{fig:resultsqcd}d. In
general, the (next-to-leading) dominance of $t\rightarrow c\,H^{0}$ 
in Figs.\,\ref{fig:maxims}a and b is confined to
a small corner of the parameter space which we pick up during the process of
maximization.
Quite in contrast, the FCNC top quark
decay into the lightest Higgs scalar can have an observable ratio in a large
portion of the parameter space, and in particular for almost all the range
of Higgs boson masses. Needless to say, not all of the maxima can be
simultaneously attained as they are obtained for different values of the
parameters.

{Moreover, we have also found ({Cf. Fig.\,\ref{fig:maxims}c)}
the maximum FCNC rate of the gluon channel in the MSSM. {Under the
RG-based assumption of only mixing in the LH sector}, it reads} 
\begin{equation}
B^{\mathrm{SUSY-QCD}}({t\rightarrow c\,g})\lsim \,10^{-5},
\label{corner2}
\end{equation}
{but it never really reaches the critical value} $10^{-5},$ {which
  can be considered as the visible threshold for the next generation of 
colliders (see below).} {{The visible limit for the gluon channel
can only be picked up if one maximizes the ratio under the assumption of
both LH and RH similar contributions, but even in this case the limit is
only barely reached}. However, we emphasize that the right order of
magnitude can already be achieved with only flavor mixing in the
LH sector. So for this decay we confirm the recent analysis of Ref.\cite
{gabbiani} in contrast to that of Ref.\,\cite{Koenig}.} In
Fig.\,\ref{fig:maxims}c we have plotted the maximum value of $%
B(t\rightarrow c\,g)$ as a function of $\delta
_{23}$ after scanning for the rest of the MSSM mass parameters within the $%
1\,TeV$ range. Interestingly enough we see that the maximum rate is not
reached for the highest possible value of the flavor mixing parameter
$\delta_{23}$, the
reason being that the physical squark masses -- obtained after diagonalizing the
$6\times6$ squark mass matrix in (flavor)$\times$(chiral) space -- are constrained
from below, so that the higher is $\delta_{23}$ the smaller is the maximally 
allowed value for
the left-right mixing in the stop sector --a result which also 
holds for the case of the Higgs channels (\ref{tch}).

The remarkable enhancements obtained for the top quark FCNC Higgs decays
(\ref{tch}) could bring them to detectable levels in large portions of the parameter
space and not just in small optimized domains.
{To assess the discovery reach of the FCNC top quark decays in the
next generation of accelerators} we take as a guide the estimations that
have been made for gauge boson final estates \cite{limits}. Using the
information mentioned in Sect.\,\ref{sec:intro} and assuming that
all the FCNC decays $t\rightarrow c\,X$ ($X=V,h$) can be treated similarly,
we roughly estimate the following sensitivities for $100\;fb^{-1}$ of
integrated luminosity: 
\begin{eqnarray}
\mathrm{\mathbf{LHC:}}B(t\rightarrow c\,X) &\gsim&5\times 10^{-5}  \nonumber
\\
\mathrm{\mathbf{LC:}}B(t\rightarrow c\,X) &\gsim&5\times 10^{-4}  \nonumber
\\
\mathrm{\mathbf{TEV33:}}B(t\rightarrow c\,X) &\gsim&5\times 10^{-3}\,\,\,.
\label{sensitiv}
\end{eqnarray}
Therefore, LHC seems to be the most suitable collider where to test this
kind of phenomena. 
The LC is limited by statistics (due to much smaller top quark
cross-section) but {in compensation} every collected event is
clear-cut. So this machine could eventually be of much help, especially if
we take into account that it could deliver $500\;fb^{-1}$ {per year}\,\cite{Miller} 
during the highest luminosity phase. In fact, a high luminosity $e^{+}e^{-}$
super-collider offers the greatest potential for high precision top quark
physics and it constitutes an ideal complement to concomitant LHC
experiments. {So, even if no SUSY particle is seen at LEP II, there is
indeed a possibility to pin down quite a few FCNC }$t\rightarrow c\,h$ {%
and/or }$t\rightarrow c\,g$ {decays at the LHC first, and later on to
perform a more detailed study at the LC. The situation with the Tevatron, as
we see, is much more gloomy, since the required FCNC rates of} $\sim 10^{-3}$
{cannot be attained unless we artificially search in some remote (unnatural)
corner of the parameter space. Hence, the Tevatron (even TeV}$33$) {seems to
be of no much help in this regard, unfortunately.}

Of course, with the range of SUSY masses that we have used in our analysis,
one could try to detect some of the sparticles directly at the
super-colliders. But this is always harder than dealing with conventional
particles, {and in any case the FCNC method is complementary to direct searches}.
Therefore, from the enormous experience gained in the study of top quarks, a
dedicated analysis of the potential few hundred FCNC top quark decays per
year at the super-collider LHC could provide a more clear identification of
physical effects beyond the SM. This is so not only because the SUSY-FCNC
rates could well be within the experimentally feasible ranges~(\ref{sensitiv}%
), as we have seen in our numerical analysis, but also because the
signatures are likely to be separable from the background. {For
instance, at high }$\tan \beta \gsim 30$ {all of the final state
Higgs bosons }$h=h^{0},H^{0},A^{0}$ {would mainly
disintegrate into }$b\overline{b}$ {pairs, }$h\rightarrow b\overline{%
b}${. These decays are also highly augmented as compared to the SM
prediction, and include large MSSM radiative corrections\,\cite{CJS}.
Although there is an interval of Higgs masses where the alternative decay }$%
h\rightarrow ZZ$ {is possible and sizeable for an SM Higgs, this
decay mode is not dominant for the SUSY Higgs bosons }$h=H^{0},A^{0}$ {in
the relevant region, and it is never kinematically accessible 
to }$h^{0}${. In this region we expect that the FCNC final states }$t%
\overline{t}\rightarrow c\,h,\,\overline{c}h$ {are to be observed
mainly as }$c\,b\,\overline{b,}\;\overline{c}b\overline{b}${, and
they should be effectively tagged through high }$p_{T}$ {charm-quark
jets and large invariant mass for the recoiling }$b\overline{b}$
{pairs. On the other hand, for moderately small }$\tan \beta \gsim 1$  
{the situation is more cumbersome since the gluino-mediated FCNC
rates for }$t\rightarrow c\,h$ {can be equally sizeable as
for high }$\tan \beta${, but in turn the decay pattern of the
neutral Higgs bosons is much more complicated, and this is so the heavier
are the Higgs bosons \cite{Djouadi}. Notwithstanding, since in our case the
Higgs bosons must satisfy }$m_{h}<m_{t}${, it turns out that both
the lightest CP-even and the CP-odd states }$h=h^{0},A^{0}$ {still
do preferentially decay into }$b\overline{b}$ {pairs. But not
necessarily so the heavy CP-even Higgs }$H^{0}${, for in this
region it first cascades down predominantly (}$90\%$) {into }$h^{0}$
$h^{0}$ ({except in a narrow interval centered around
  }$M_{H^{0}}=130\,GeV$ {where }$H^{0}\rightarrow b\overline{b}$
{again dominates because the }$H^{0}h^{0}h^{0}$-{coupling changes
  sign}){; and as a result the observable }$b\overline{b}$ {pairs
  in 
the final state  (emerging from the two secondary -- real or virtual -- decays }
$h^{0}\rightarrow b\overline{b}$) {are much softer than in the previous situation.
Fortunately, we have seen that among the FCNC decays }$t\rightarrow c\,h$  
{the one corresponding to }$h=H^{0}$ {is typically the most
suppressed one, except in some corners of parameter space. 
In contrast, the most relevant decay (both under typical
and optimized conditions) is  that of }$h=h^{0}${. The corresponding channel is
always kinematically open in the MSSM and it can be tagged through the final
state signature }$c\,b\,\overline{b}$ {mentioned above, which is
dominant for any }$\tan \beta >1$. As for the FCNC decays in the
gluon channel, $t\rightarrow c\,g$, here also a high $p_{T}$ charm-quark jet
against a highly energetic ($\sim m_{t}/2$) gluon jet is the indelible
imprint of this decay. Altogether these features should be very helpful to
mostly reduce all backgrounds and provide a clear-cut signal of rare top
quark decays beyond the SM.

To conclude, the FCNC decays of the top quark are such rare events in the SM
(especially the top quark decay into the Higgs boson) that their observation
at detectable levels should be interpreted as an extremely robust indication
of new physics. At present the most popular (phenomenologically
compatible\,\cite{HollWeig} and technically fully consistent\,\cite{MSSM})
quantum field 
theoretical extension of the SM is the MSSM; therefore, the effective
tagging of FCNC top quark decays in the major accelerators round the corner
could well be a first step into discovering SUSY. From our analysis we find
that around the loci of maxima of the rates for the leading FCNC top quark decays,
$t\rightarrow c\,h$ and $t\rightarrow c\,g$, we roughly have the
following situation (under the assumption of flavor mixing only
in the LH sector): } 
\begin{equation}
5\times 10^{-6}\lsim B(t\rightarrow c\,g)_{\max }<B(t\rightarrow
c\,h)_{\max }\lsim 5\times 10^{-4}.  \label{maxs}
\end{equation}
{In both types of decays the dominant effects come from SUSY-QCD.
However, it should not be undervalued the fact that the maximum electroweak
rates for }$t\rightarrow c\,h$ {can reach the }$10^{-6}$ {level, which amounts 
not only to saying that they are }$7$ {orders
of magnitude greater than the maximum SM ones, but also to noticing that
they are on the verge of being detectable. Last but not least, we stress
once again that the largest FCNC rate both from SUSY-QCD and SUSY-EW is
precisely that of the lightest CP-even state}, {a very fortunate
fact which can hardly be overemphasized as  }$t\rightarrow c\,h^{0}$ 
{is the only Higgs channel that is phase-space available across the
whole MSSM parameter space.}

{At the end of the day, we should seriously bear in mind the possibility of
observing FCNC} {decays of the top quark, and so we ought to be prepared to
collect a few hundred, perhaps a few thousand, events of this sort at the
LHC, together with a few dozen (although crystal-clear) }$t\rightarrow c\,X$
($X=h,g$) {decays} at the LC. If that would be the case, we could eventually
find ourselves
in the process of discovering SUSY dynamics at work before being able to directly
produce (or clearly identify) any supersymmetric particle.

\textbf{Acknowledgments:} One of the authors (J.S.) is thankful to W.
Bernreuther and D. Miller for updated information on the LC program. We are
also grateful to M. Mart{{\'\i}}nez for similar information. {The work
of J.G}. has been supported in part by a grant of the Comissionat per a
Universitats i Recerca, Generalitat de Catalunya No. FI95-2125. This work
has also been financed by CICYT under project No. AEN95-0882.


\newpage

\section*{Figure Captions}

\renewcommand{\labelenumi}{{\bf Fig.~\theenumi}}

\begin{enumerate}
\item  One-loop SUSY-EW vertex diagrams for the decay $\tch$
  $(h=h^{0},H^{0},A^{0})$. Here $d$ ($\sd_{\{a,b\}}$) represent mass-eigenstate
  down type quarks (squarks) of any generation\label{diag:fcncEWvertex}.

\item  One-loop SUSY-EW diagrams contributing to the mixed $t-c$
self-energy, with a notation similar to that of Fig.\,\ref{diag:fcncEWvertex}.\label{diag:fcncEWmix}

\item  {One-loop SUSY-QCD diagrams for the decay }$\tch$: \textbf{(a)%
} vertex diagram, \textbf{(b)} mixed $t-c$ self-energy. $\su_{\{\alpha
,\beta \}}$ stand for mass-eigenstate up-type squarks of any generation\label%
{diag:fcncQCDvertex}.

\item  Evolution {of the SUSY-EW contributions to the ratio} (\ref
{eq:defbr}) with \textbf{(a)} ${\tan \beta }$, \textbf{(b)} the trilinear
coupling $A_{b}$, \textbf{(c)} the higgsino mass parameter $\mu $, and 
\textbf{(d)} the pseudoscalar Higgs mass $M_{A^{0}}$. The rest of inputs are
given in eq.\,(\ref{eq:inputew}). \label{fig:resultsew}

\item  Evolution of the {SUSY-QCD contributions to the ratio} (\ref
{eq:defbr}) with \textbf{(a)} the mixing parameter $\delta _{23}$ between
the 2nd and 3rd squark generations, \textbf{(b)} the higgsino mass parameter 
$\mu $, \textbf{(c)} the gluino mass $m_{\tilde{g}}$, and \textbf{(d)} the
pseudoscalar Higgs mass $M_{A^{0}}$. The rest of inputs are given in
eqs.({\ref{eq:inputew}}) and~({\ref{eq:inputqcd}}).\label{fig:resultsqcd} 

\item  Evolution of the {SUSY-QCD contributions to the ratio} (\ref
{eq:defbr}) with \textbf{(a)} the lightest sbottom mass ($\msbo$), \textbf{(b)} the lightest stop mass (\msto), and \textbf{(c)}
$\tb$.\label{fig:resultsqcdmasses}

\item  Evolution {of the SUSY-QCD effects} on $B(t\rightarrow c\,g)$
as a function of \textbf{(a)} the mixing parameter $\delta _{23}$, and 
\textbf{(b)} the gluino mass $\mg$. The results are shown under the
assumptions of: mixing only in the left-handed squark sector (solid); and
equal mixing in the left- and right-handed squark sectors
(dashed).\label{fig:tcgd23mg} 

\item  As in Fig.\,\ref{fig:tcgd23mg} but as a function of \textbf{(a)} the
  higgsino mass parameter $\mu $, and \textbf{(b)} $\tb$.\label{fig:tcgmutb} 

\item  As in Fig.\,\ref{fig:tcgd23mg} but as a function of \textbf{(a)} the
  lightest stop mass $\msto$ and \textbf{(b)} the soft SUSY-breaking trilinear
  top-squark coupling $A_{t}$. \label{fig:tcgmstAt}

\item  \textbf{(a) }Maximum value of $B(t\rightarrow c\,h)$, obtained by
taking into account only the SUSY-EW contributions, as a function of $\mA$ ; 
\textbf{(b)} as in (a) but taking into account only the SUSY-QCD
contributions; and \textbf{(c)} maximum value of $B(t\rightarrow c\,g)$ as a
function of the intergenerational mixing parameter $\delta _{23}$ in the LH
sector. In all cases the scanning for the rest of parameters of the MSSM has
been performed within the phenomenologically allowed region.\label{fig:maxims}
\end{enumerate}

\renewcommand{\labelitemi}{\theenumi. } \newpage \pagestyle{empty}

\begin{center}
\includegraphics{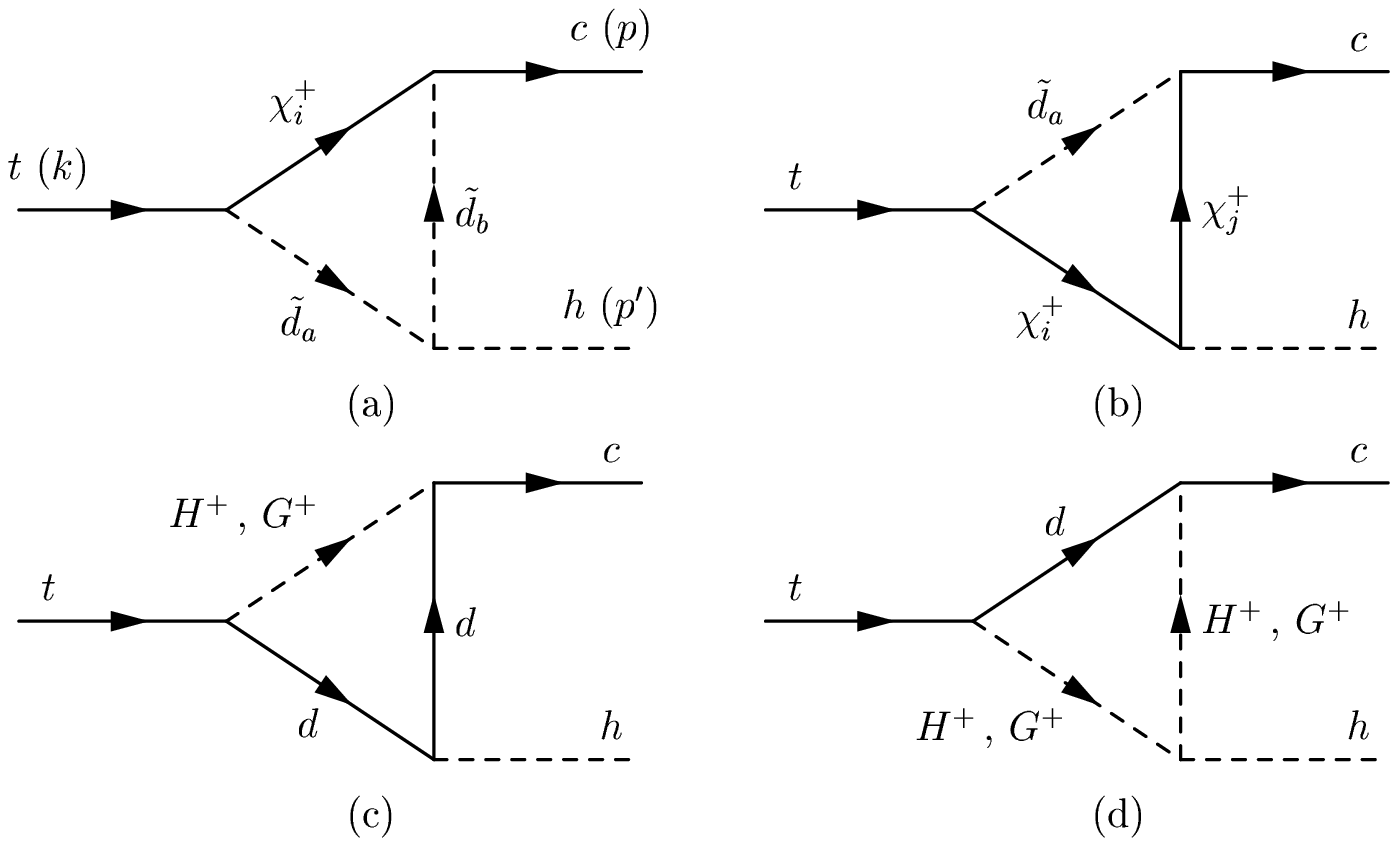}

\vspace{.3cm}

{\Large Fig. 1}
\end{center}

\vspace{1cm}

\begin{center}
\includegraphics{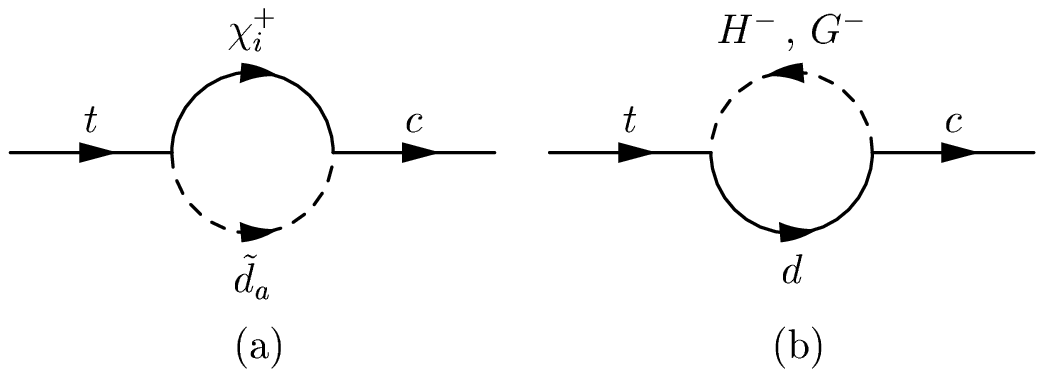}

\vspace{.3cm}

{\Large Fig. 2}

\vspace{1cm}

\includegraphics{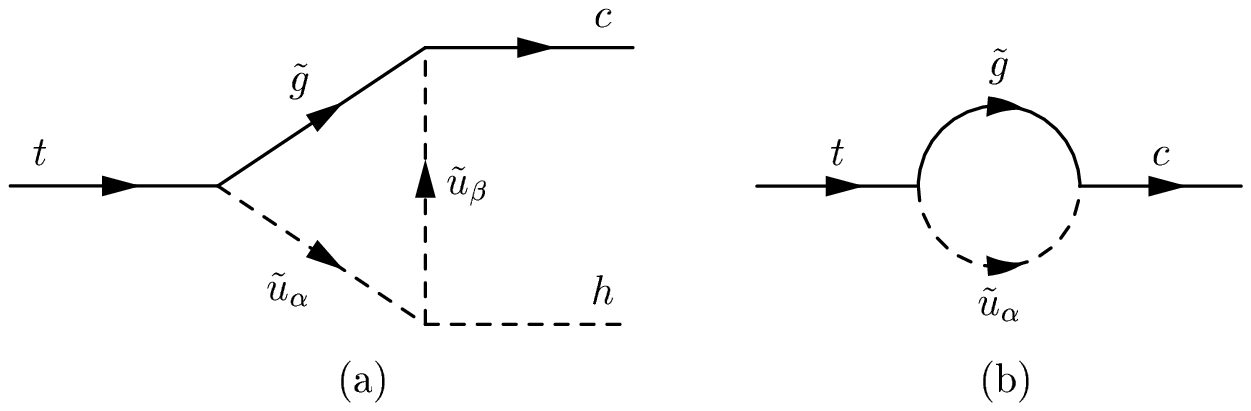}

\vspace{.3cm}

{\Large Fig. 3}



\vspace{1cm}

\begin{tabular}{cc}
\resizebox{7cm}{!}{\includegraphics{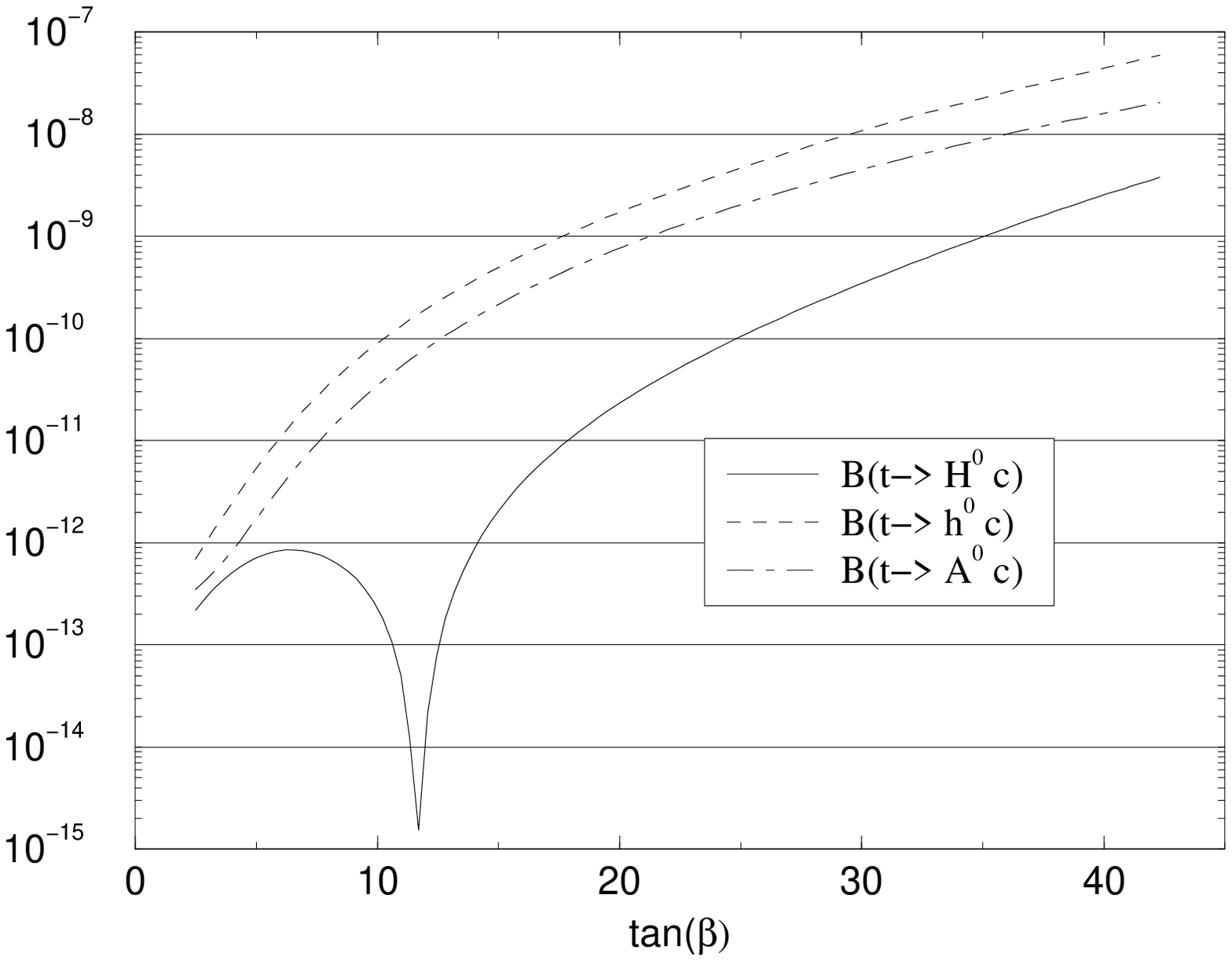}} & \resizebox{7cm}{!}{%
\includegraphics{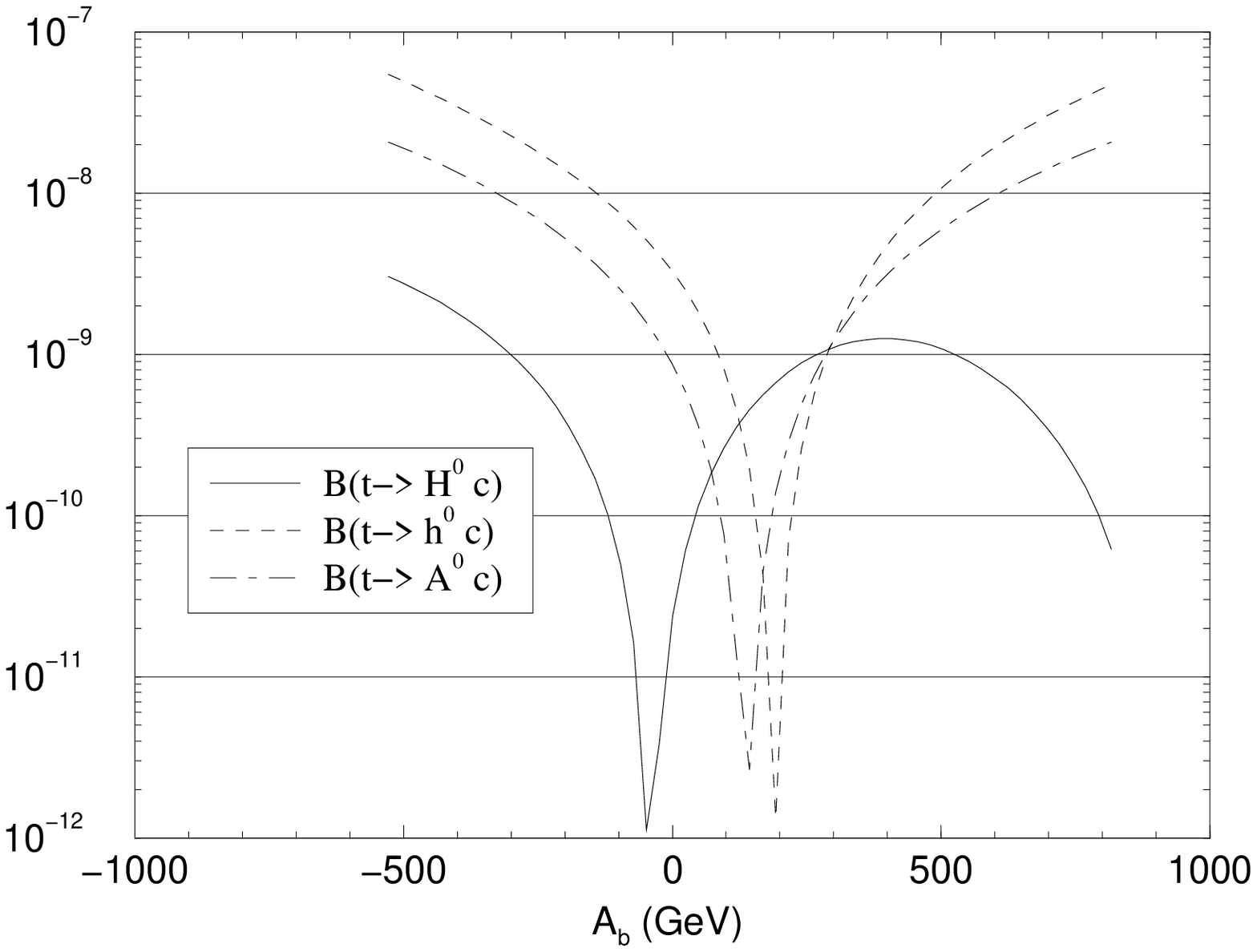}} \\ 
(a) & (b) \\ 
\resizebox{7cm}{!}{\includegraphics{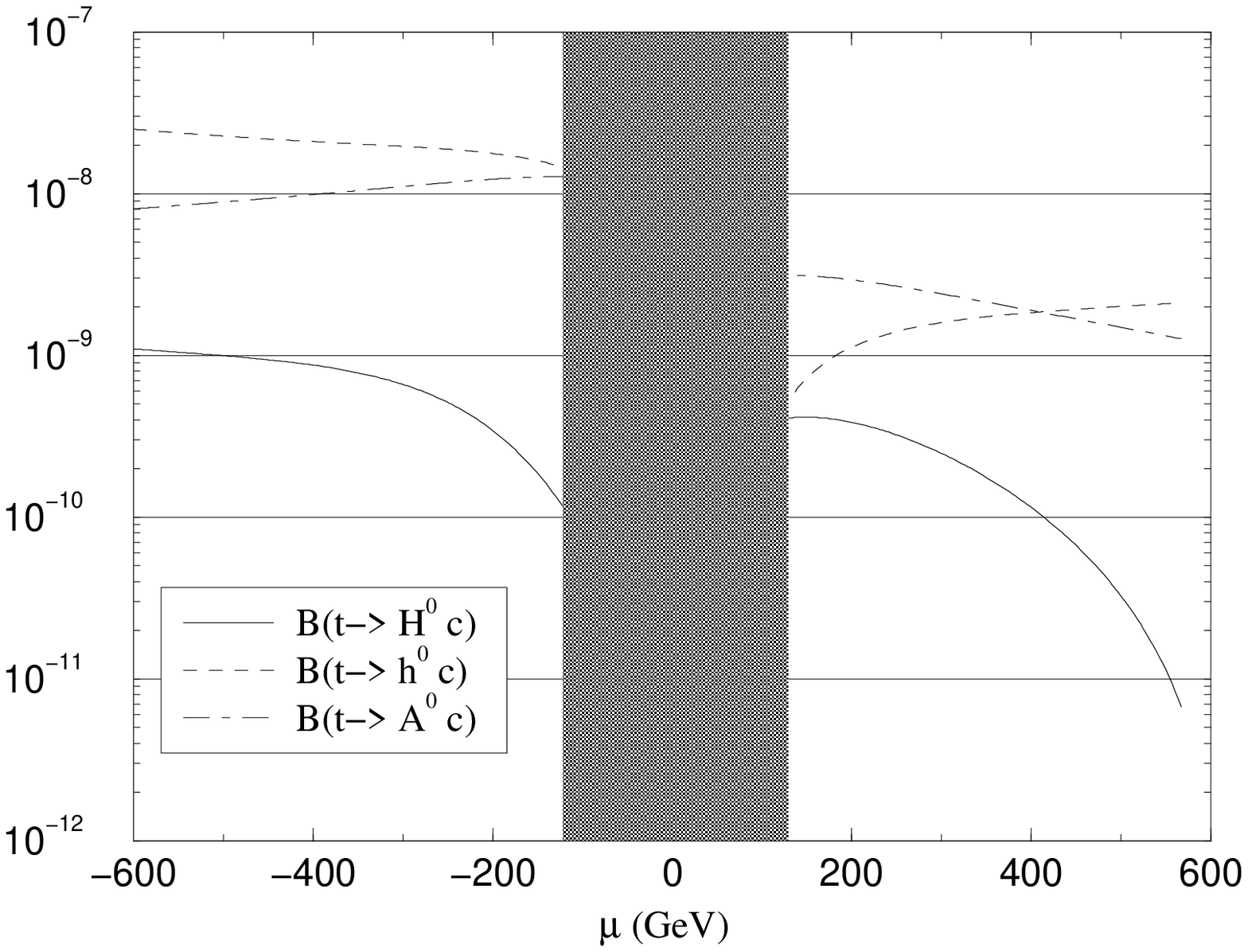}} & \resizebox{7cm}{!}{%
\includegraphics{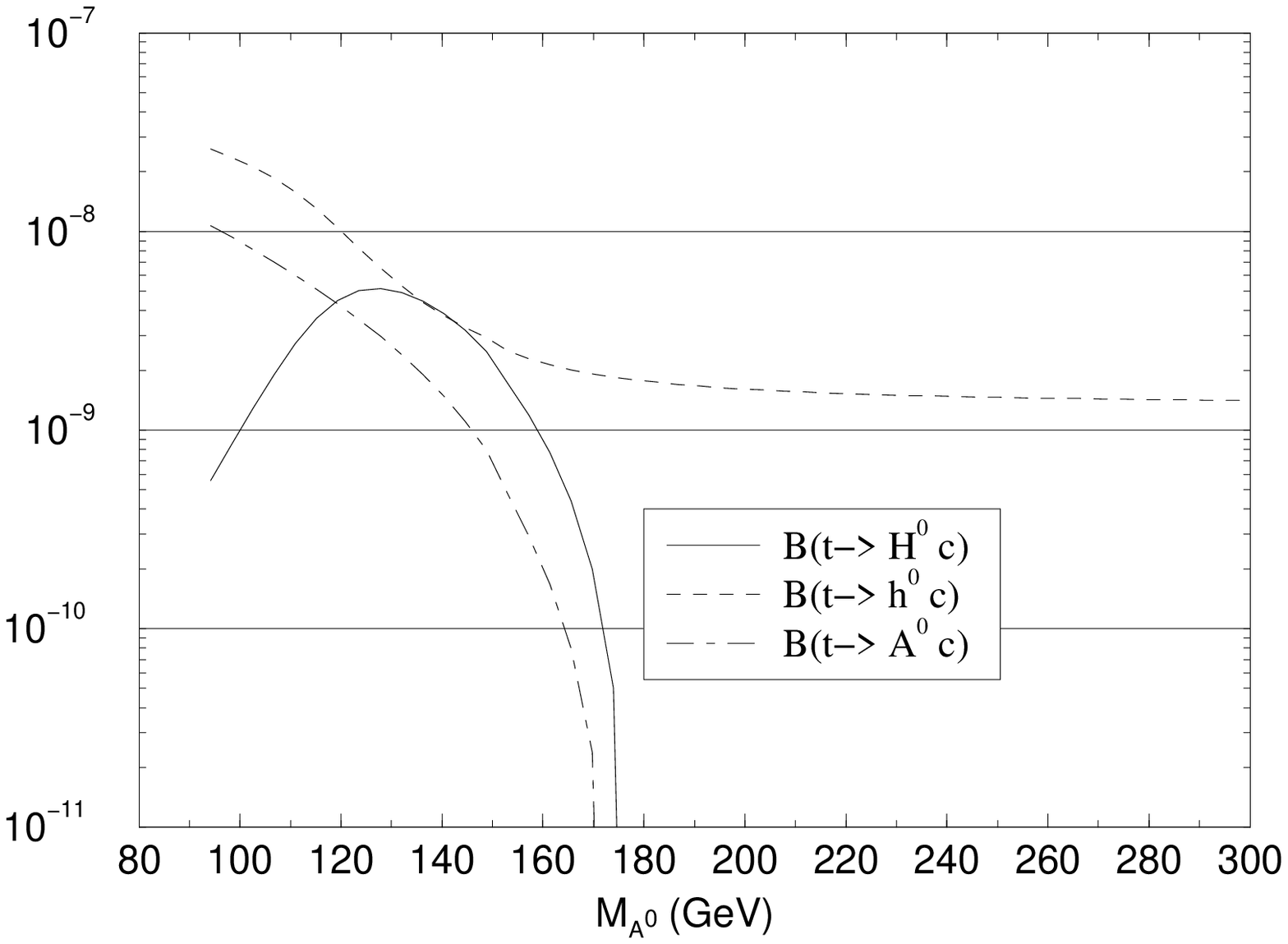}} \\ 
(c) & (d)
\end{tabular}

\vspace{.3cm}

{\Large Fig. 4}
\end{center}

\newpage

\begin{center}
\begin{tabular}{cc}
\resizebox{7cm}{!}{\includegraphics{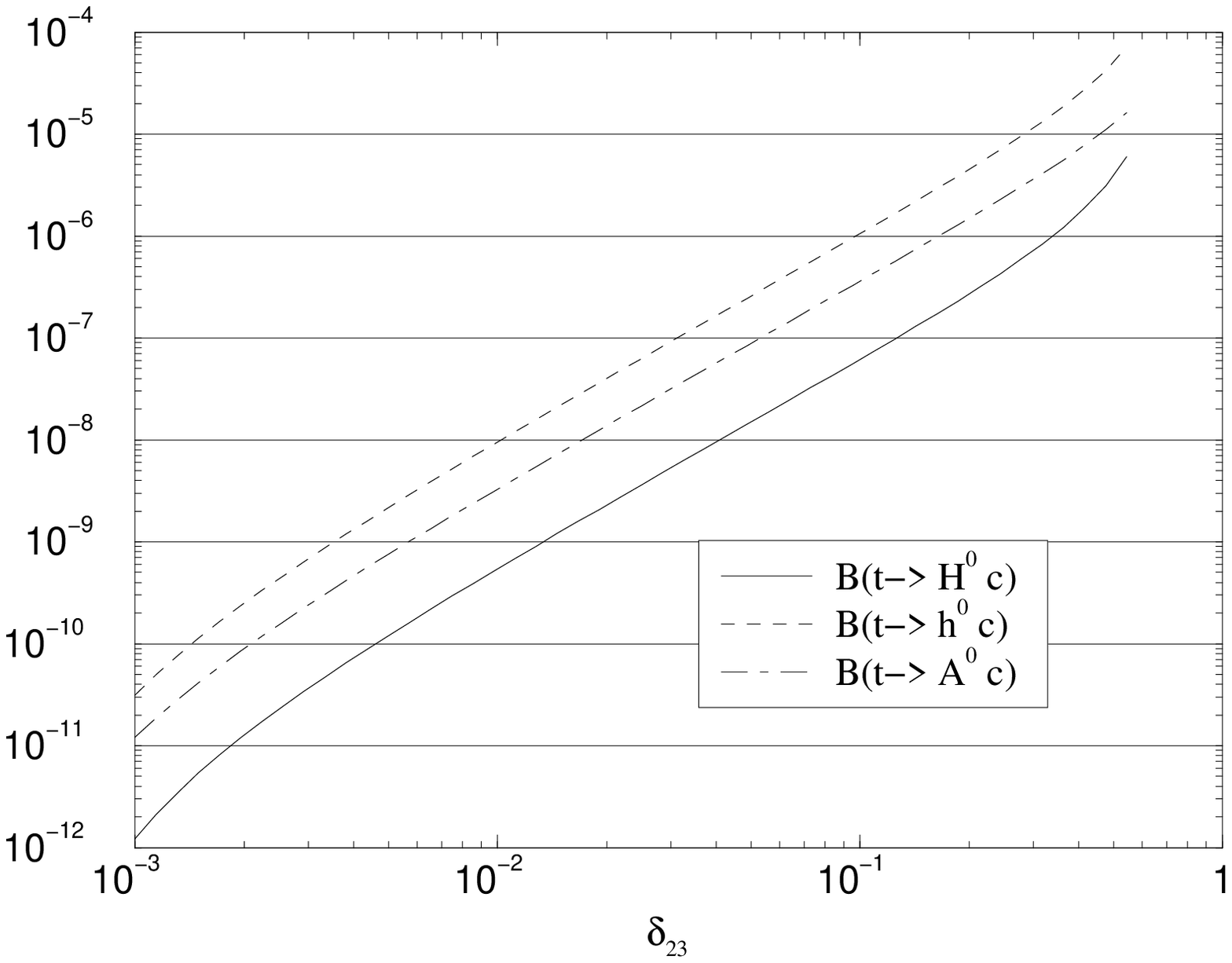}} & \resizebox{7cm}{!}{%
\includegraphics{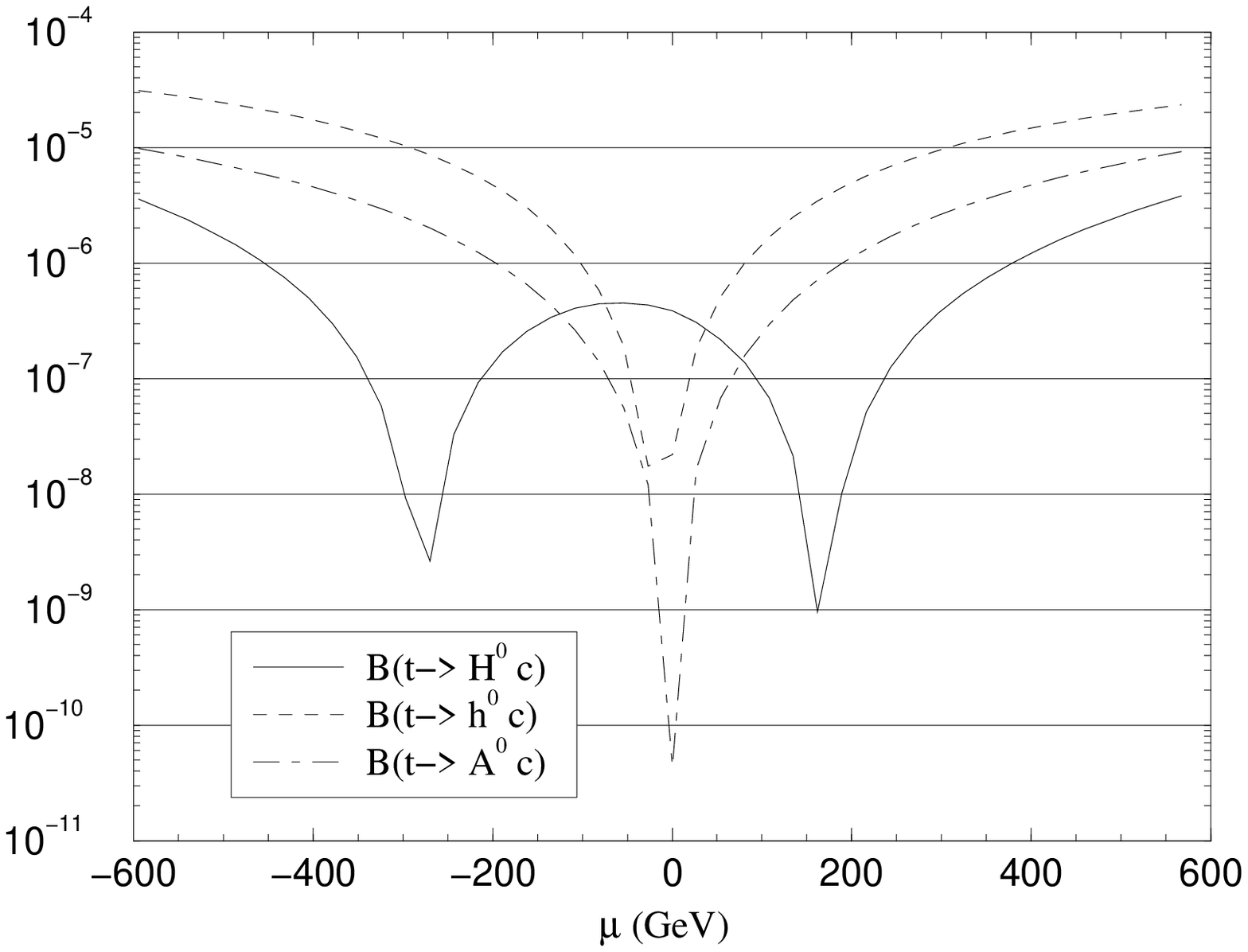}} \\ 
(a) & (b) \\ 
\resizebox{7cm}{!}{\includegraphics{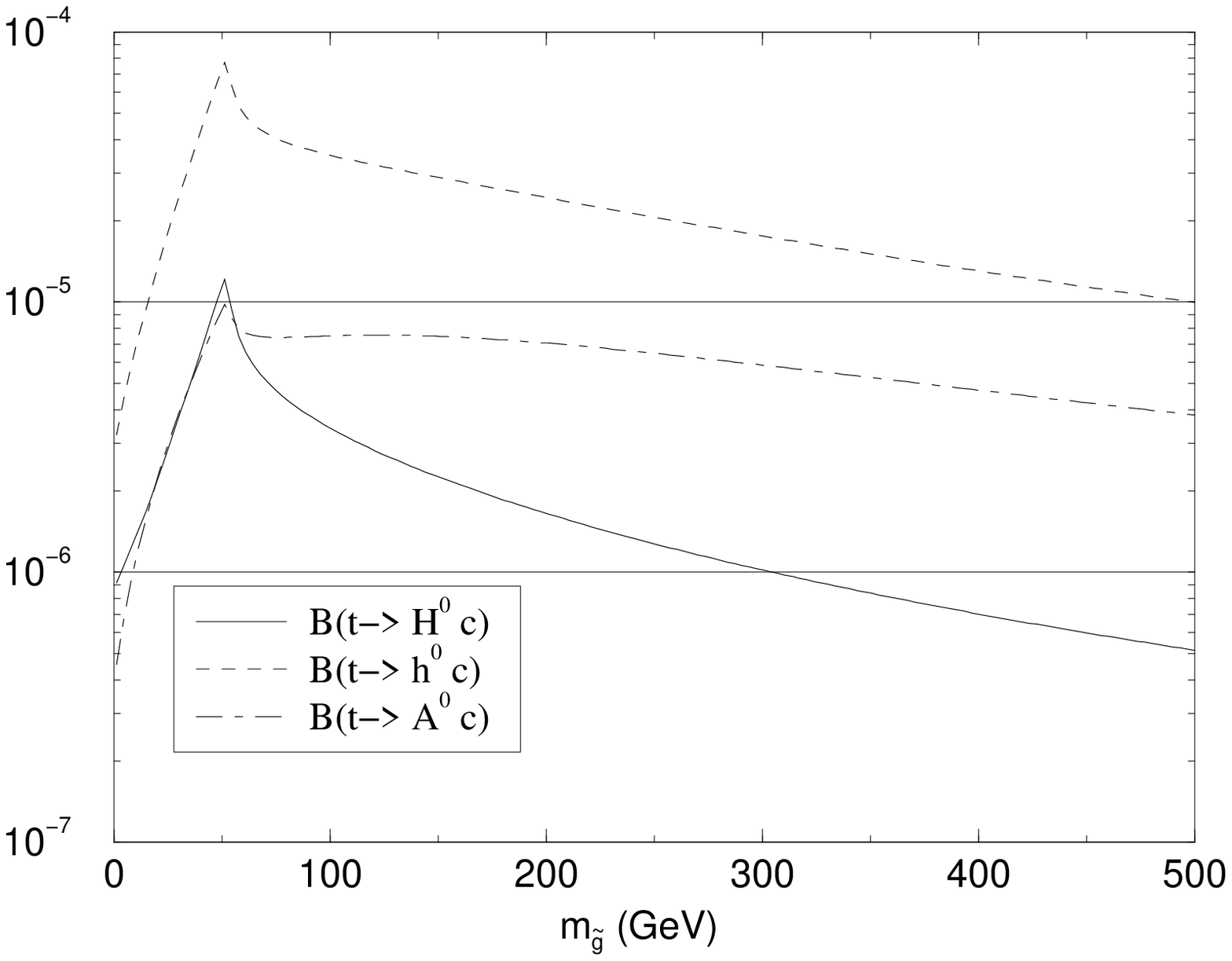}} & \resizebox{7cm}{!}{%
\includegraphics{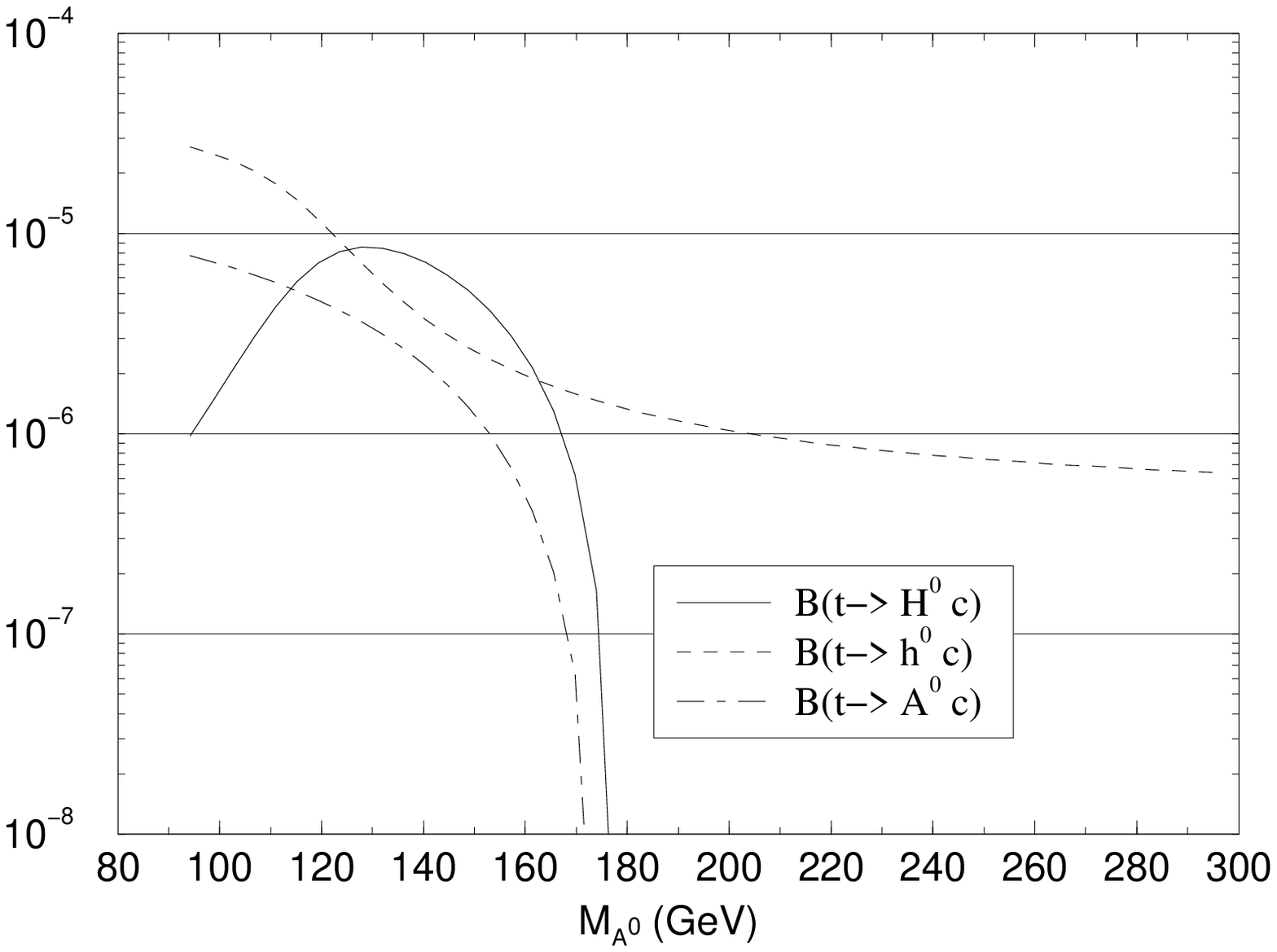}} \\ 
(c) & (d)
\end{tabular}

\vspace{.3cm}

{\Large Fig. 5}

\begin{tabular}{cc}
\resizebox{7cm}{!}{\includegraphics{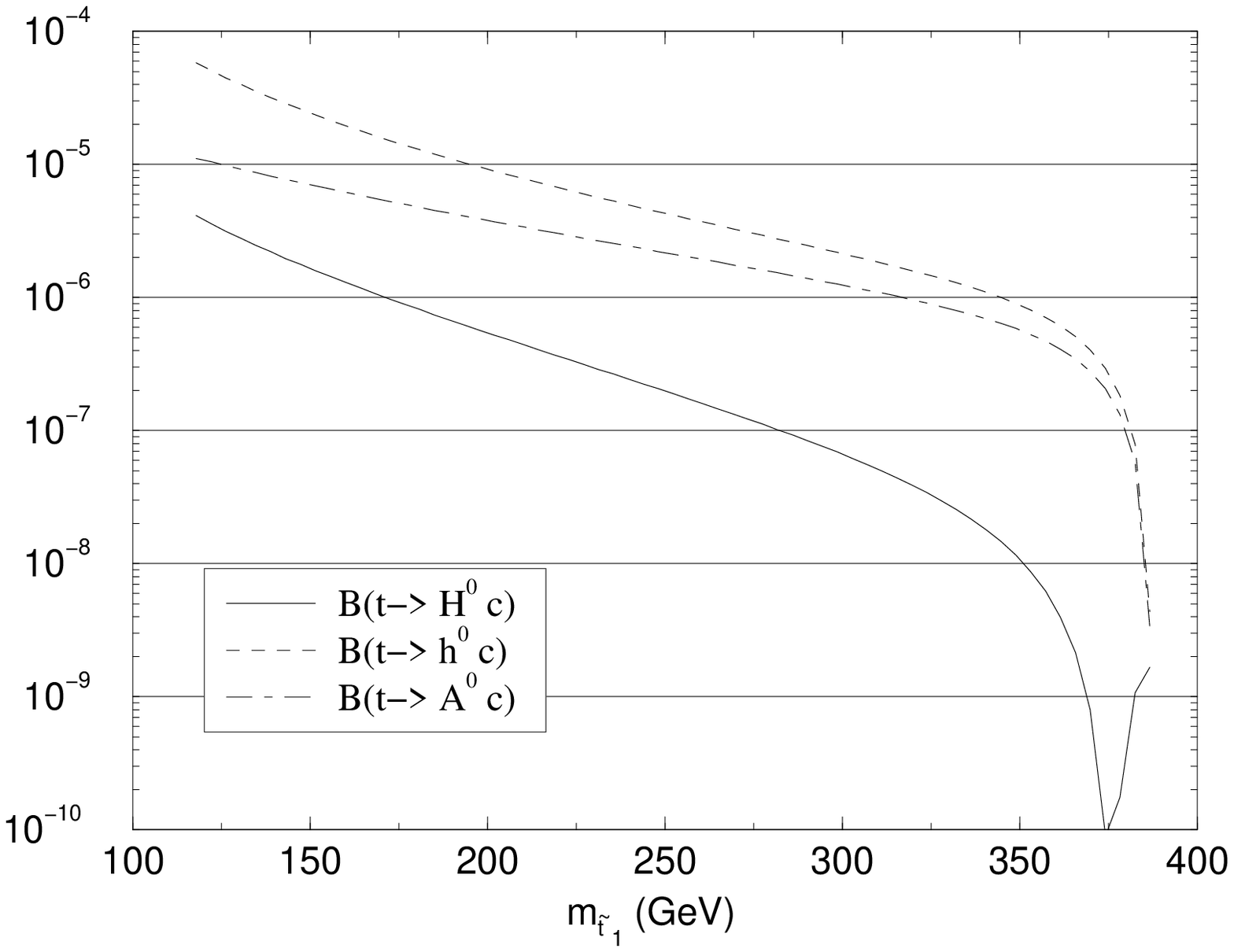}} & \resizebox{7cm}{!}{%
\includegraphics{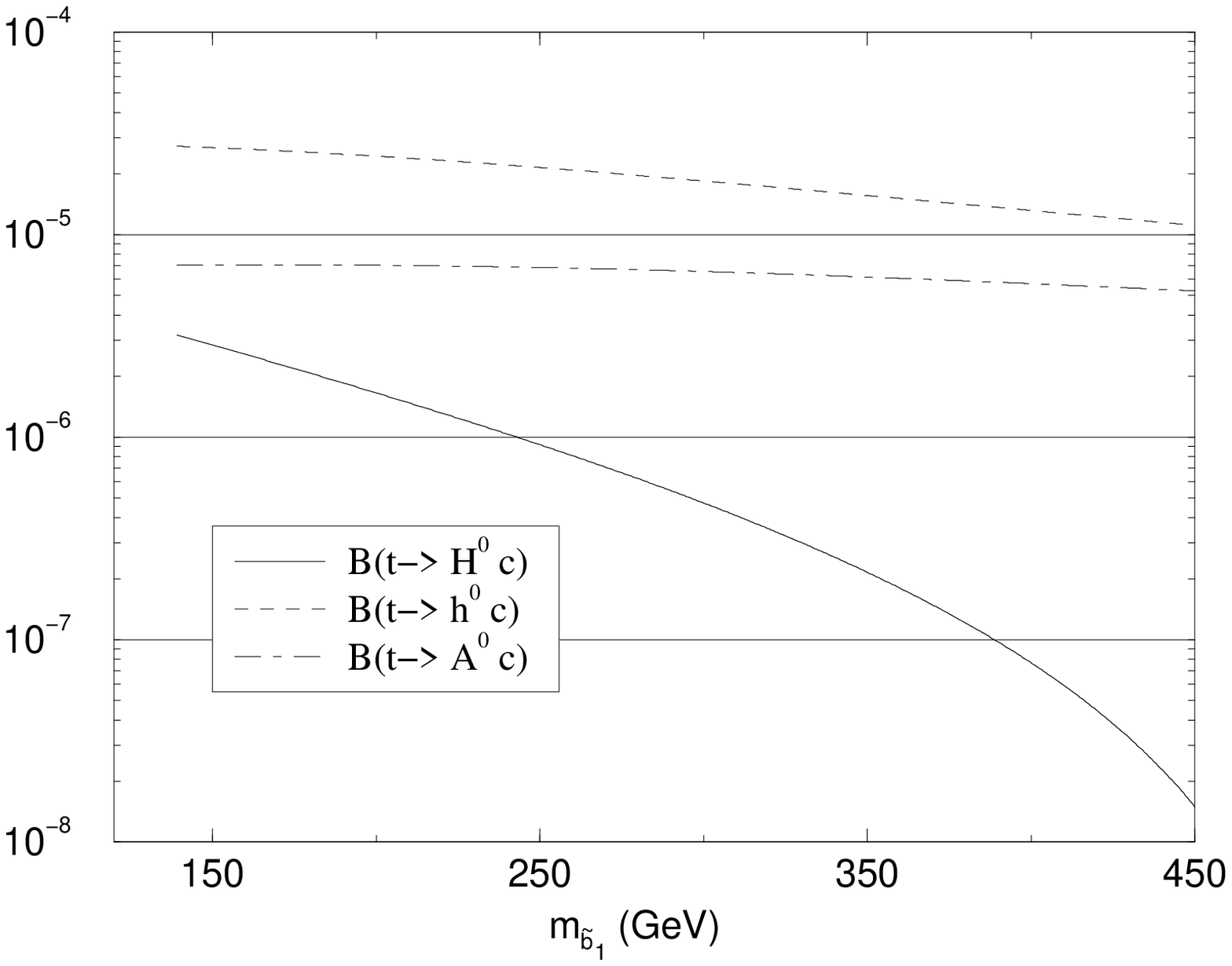}} \\ 
(a) & (b) \\ 
\multicolumn{2}{c}{\resizebox{7cm}{!}{\includegraphics{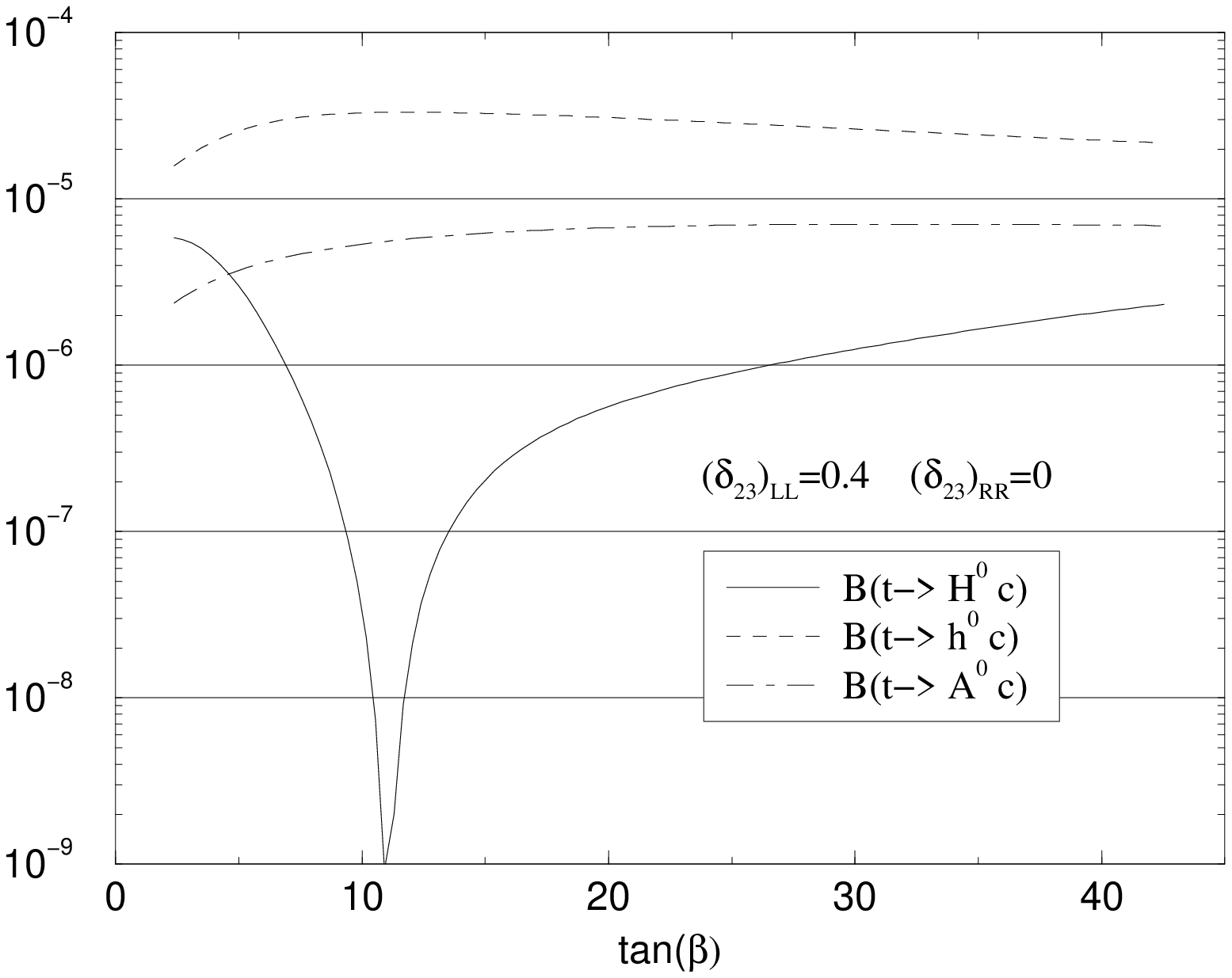}}} \\ 
\multicolumn{2}{c}{(c)}
\end{tabular}

\vspace{.3cm}

{\Large Fig. 6}
\vspace{1cm}

\begin{tabular}{cc}
\resizebox{7cm}{!}{\includegraphics{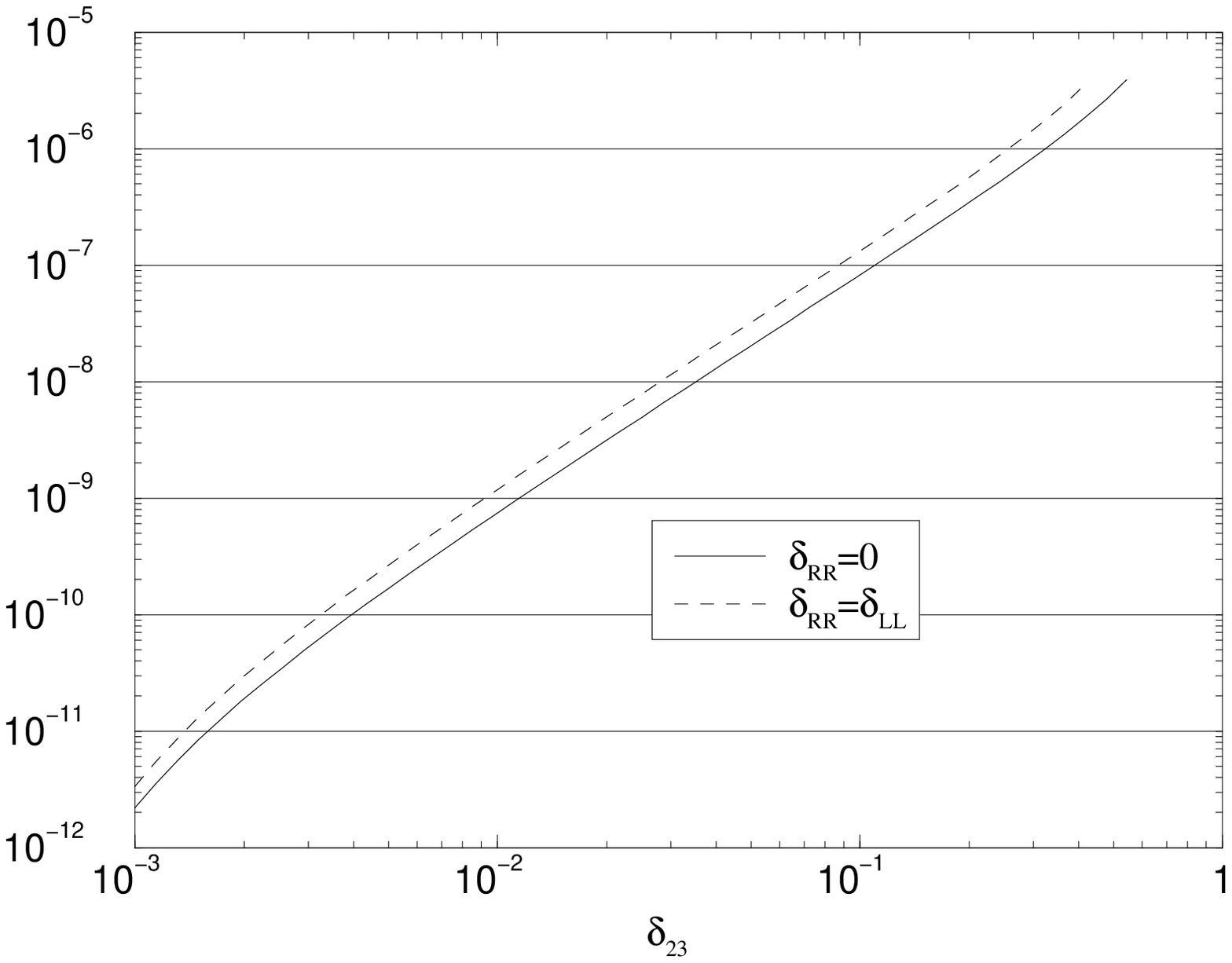}} & \resizebox{7cm}{!}{%
\includegraphics{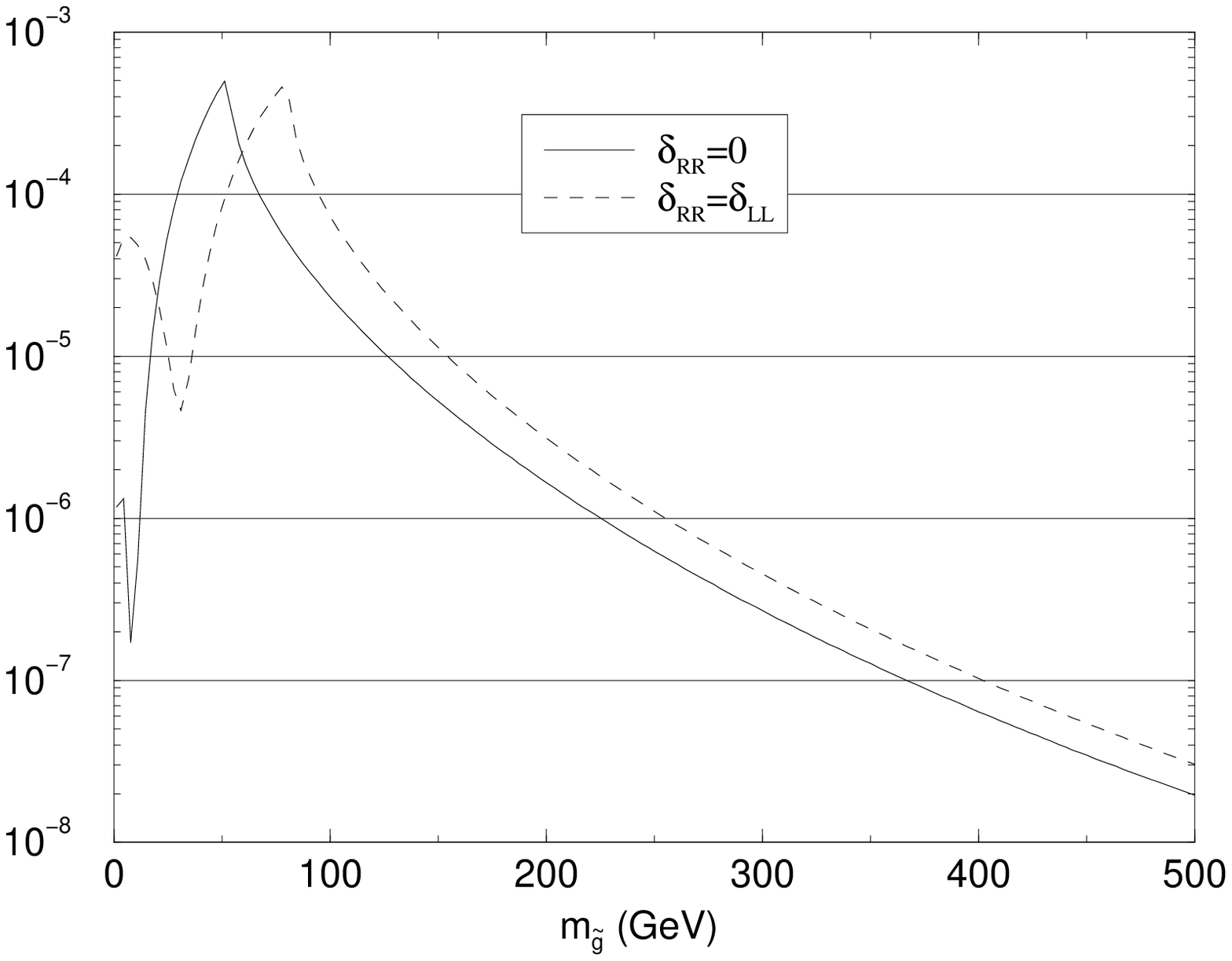}} \\ 
(a) & (b)
\end{tabular}

\vspace{.3cm}

{\Large Fig. 7}

\begin{tabular}{cc}
\resizebox{7cm}{!}{\includegraphics{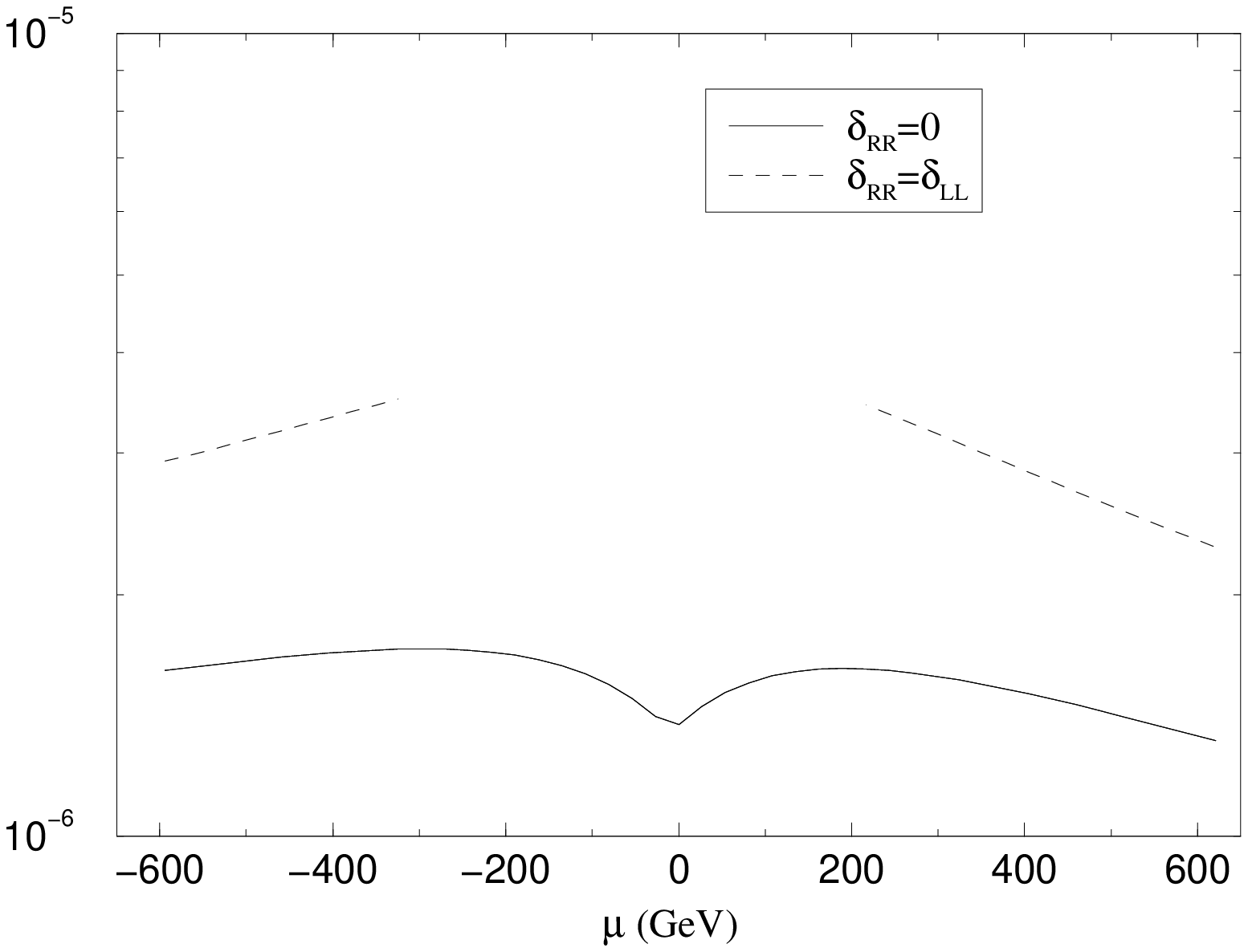}} & \resizebox{7cm}{!}{%
\includegraphics{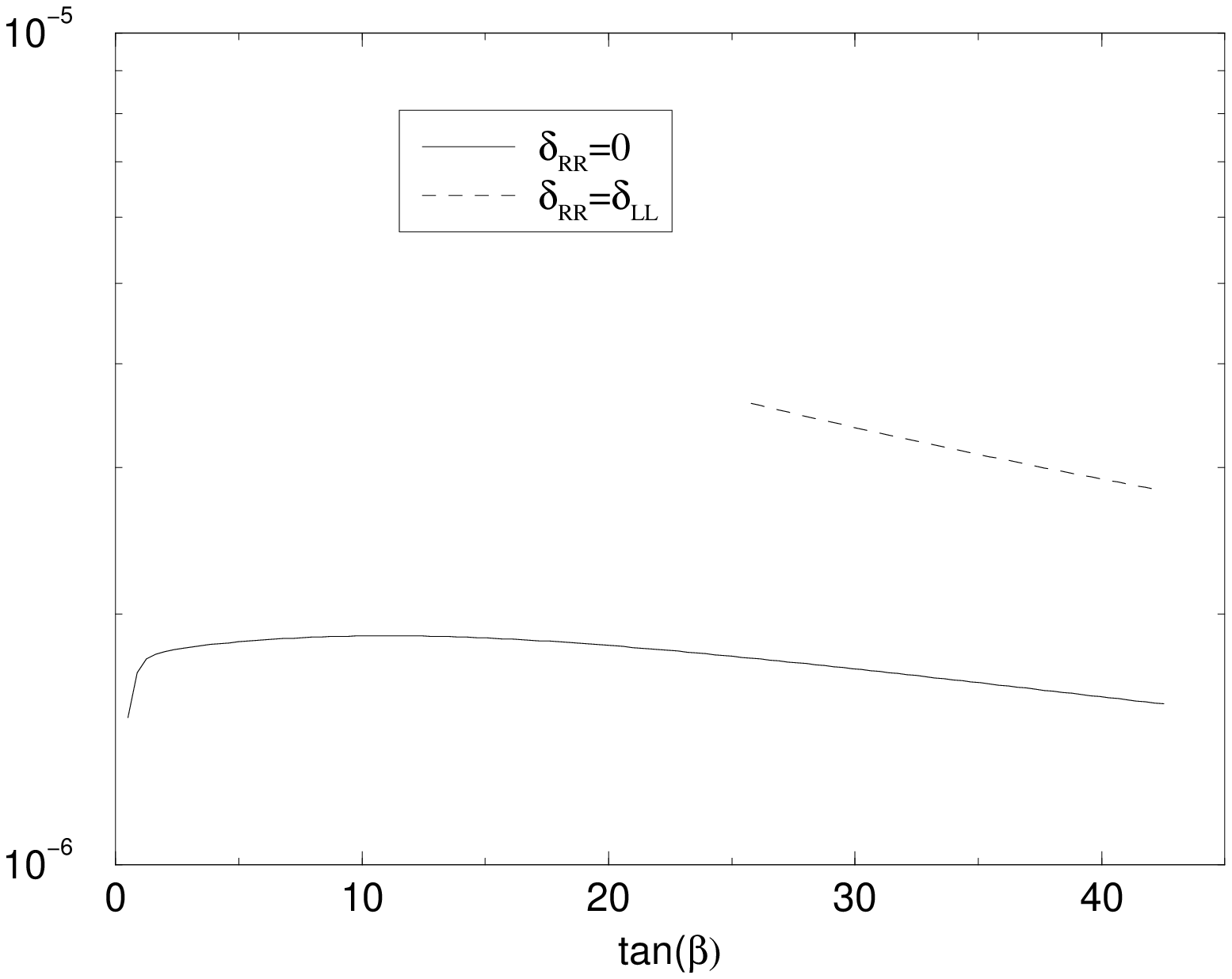}} \\ 
(a) & (b)
\end{tabular}

\vspace{.3cm}

{\Large Fig. 8}
\vspace{1cm}

\begin{tabular}{cc}
\resizebox{7cm}{!}{\includegraphics{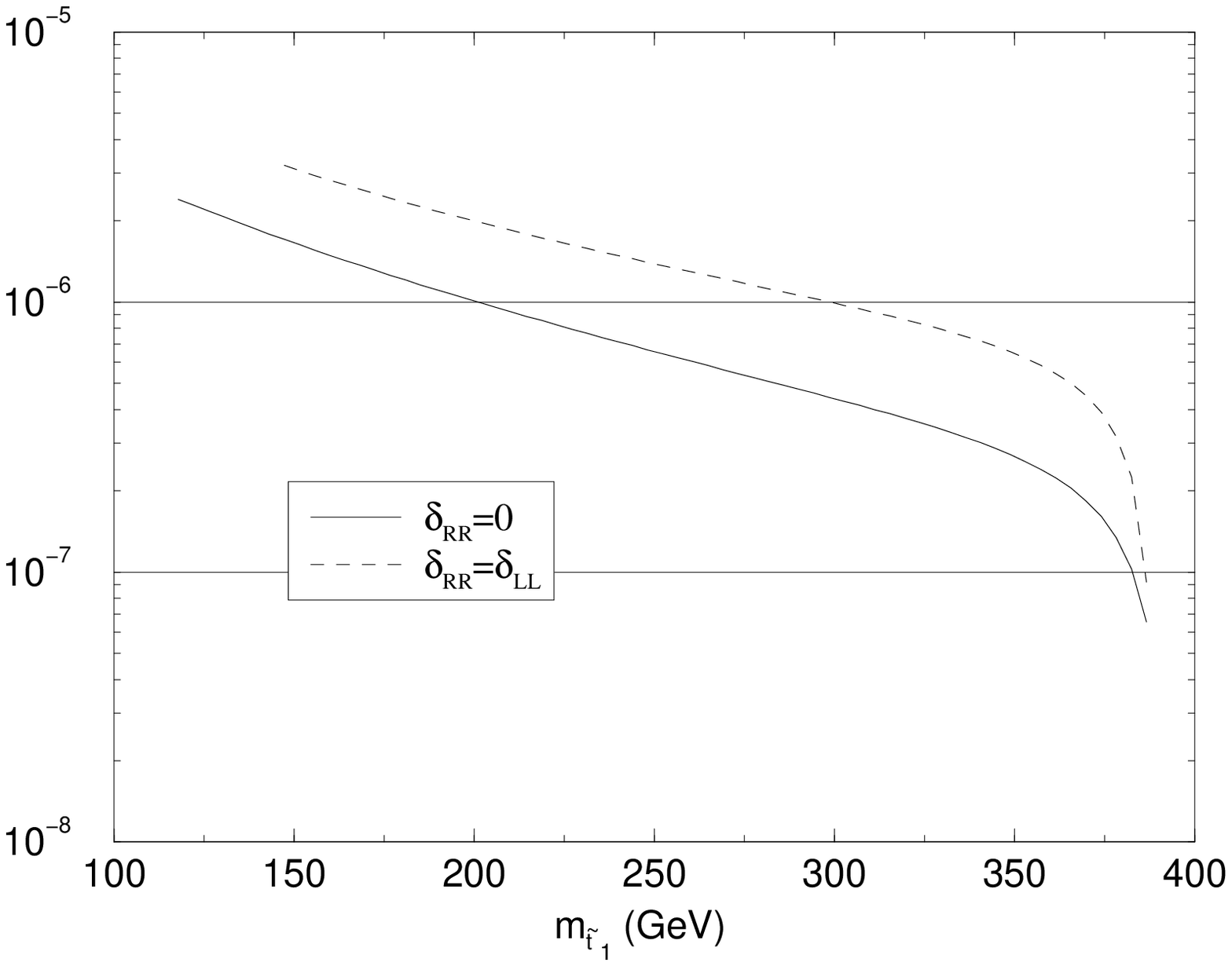}} & \resizebox{7cm}{!}{%
\includegraphics{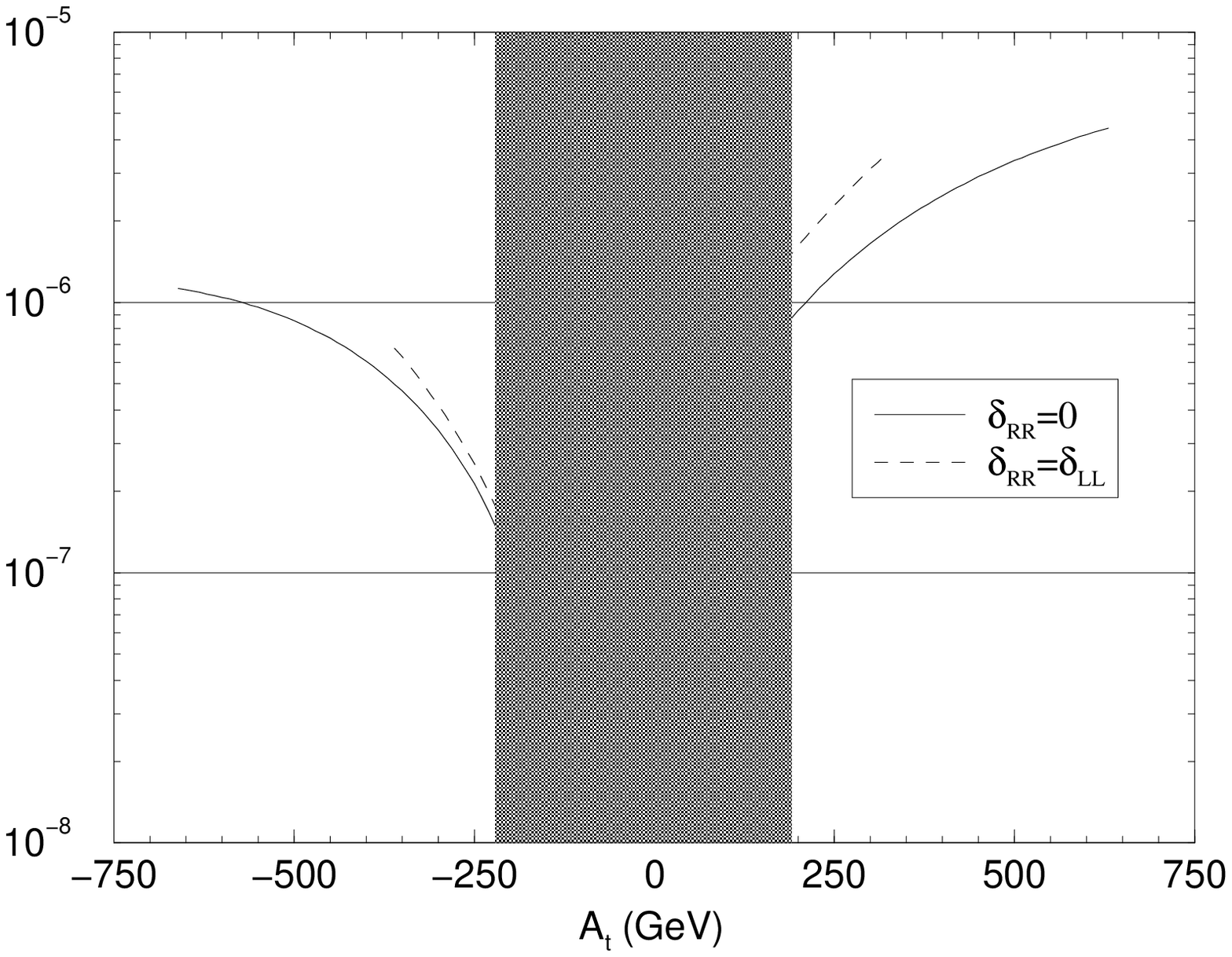}} \\ 
(a) & (b)
\end{tabular}

\vspace{.3cm}

{\Large Fig. 9}

\begin{tabular}{cc}
\resizebox{7cm}{!}{\includegraphics{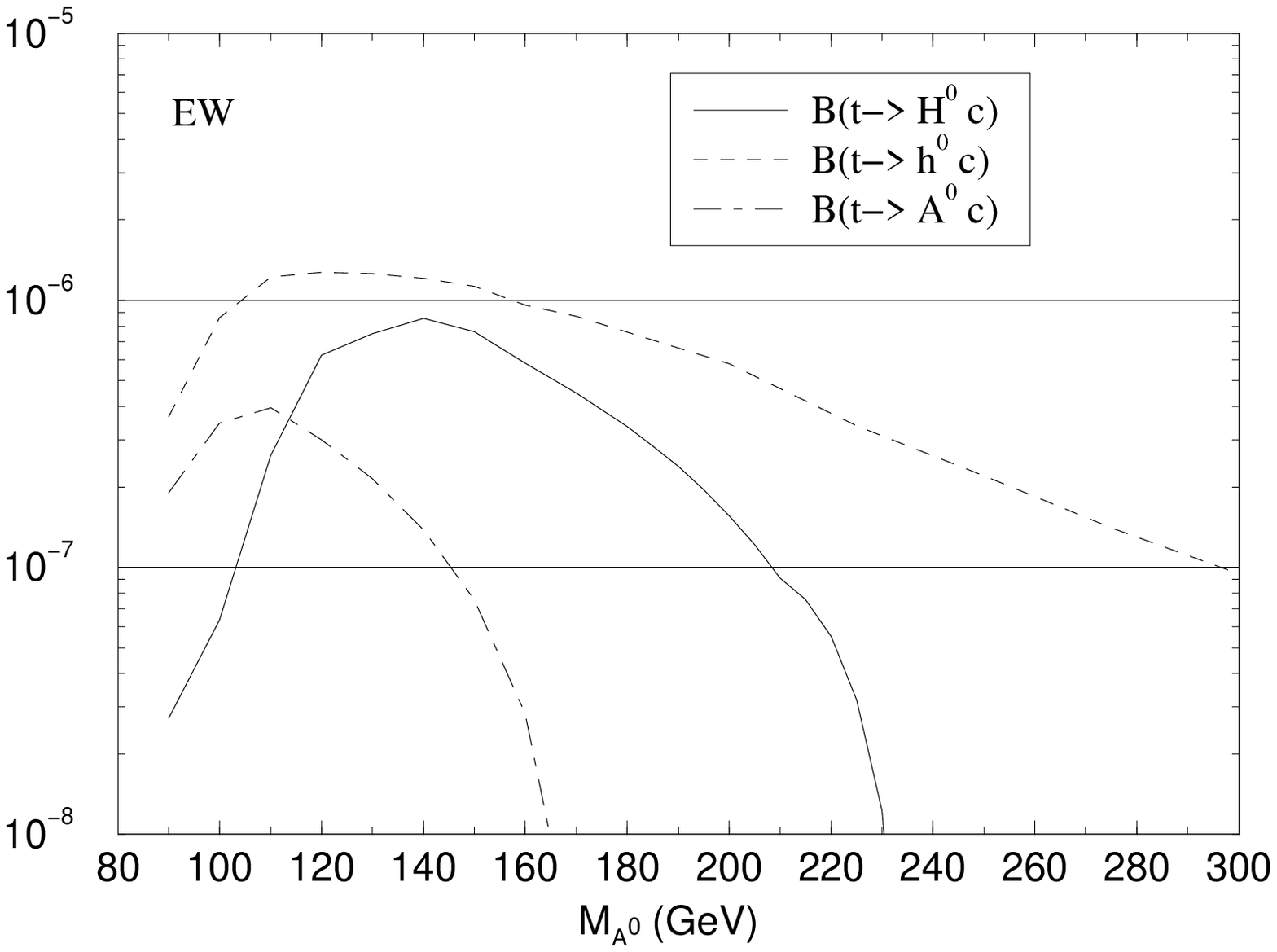}} & \resizebox{7cm}{!}{%
\includegraphics{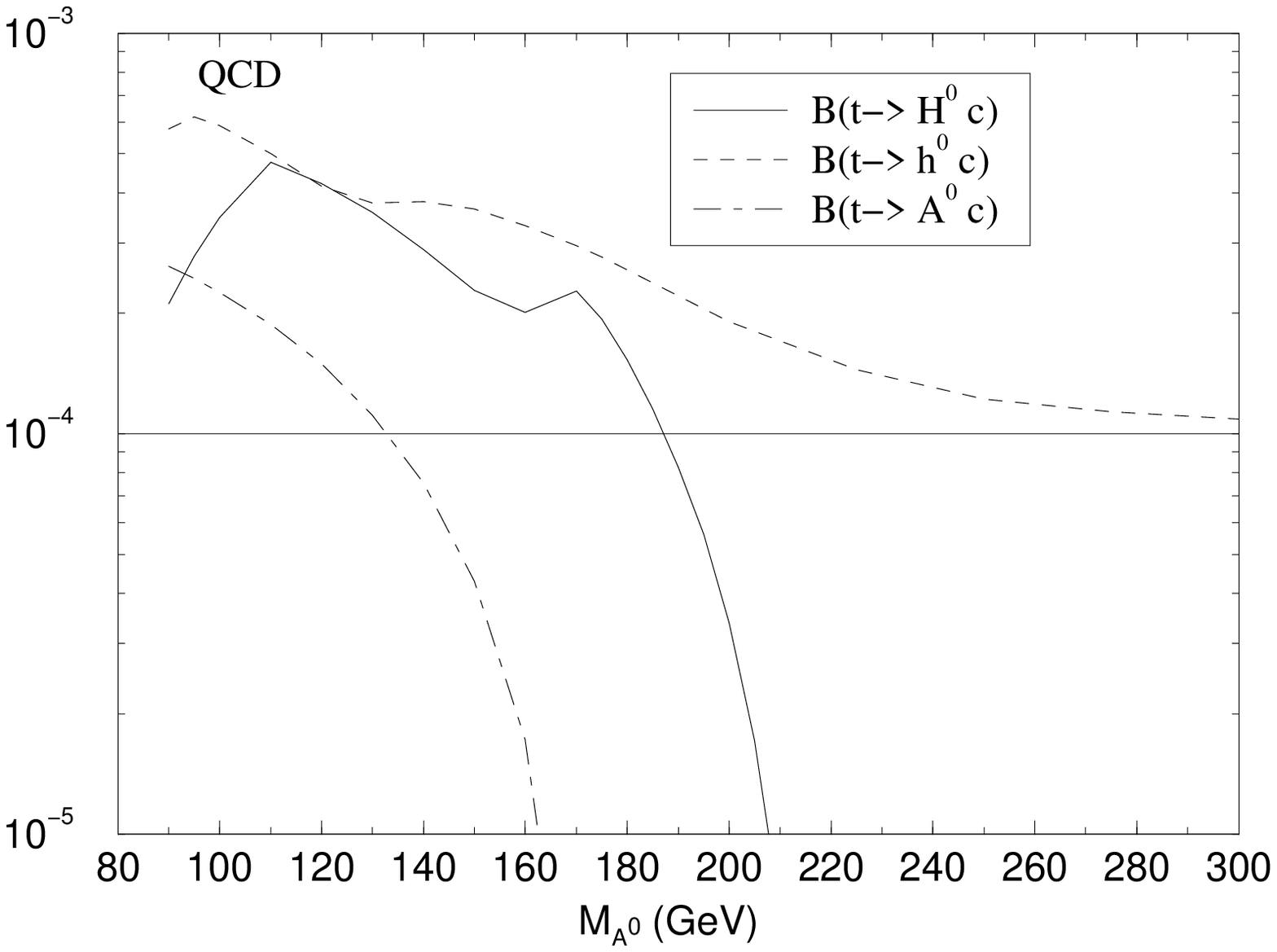}} \\ 
(a) & (b) \\ 
\multicolumn{2}{c}{\resizebox{7cm}{!}{\includegraphics{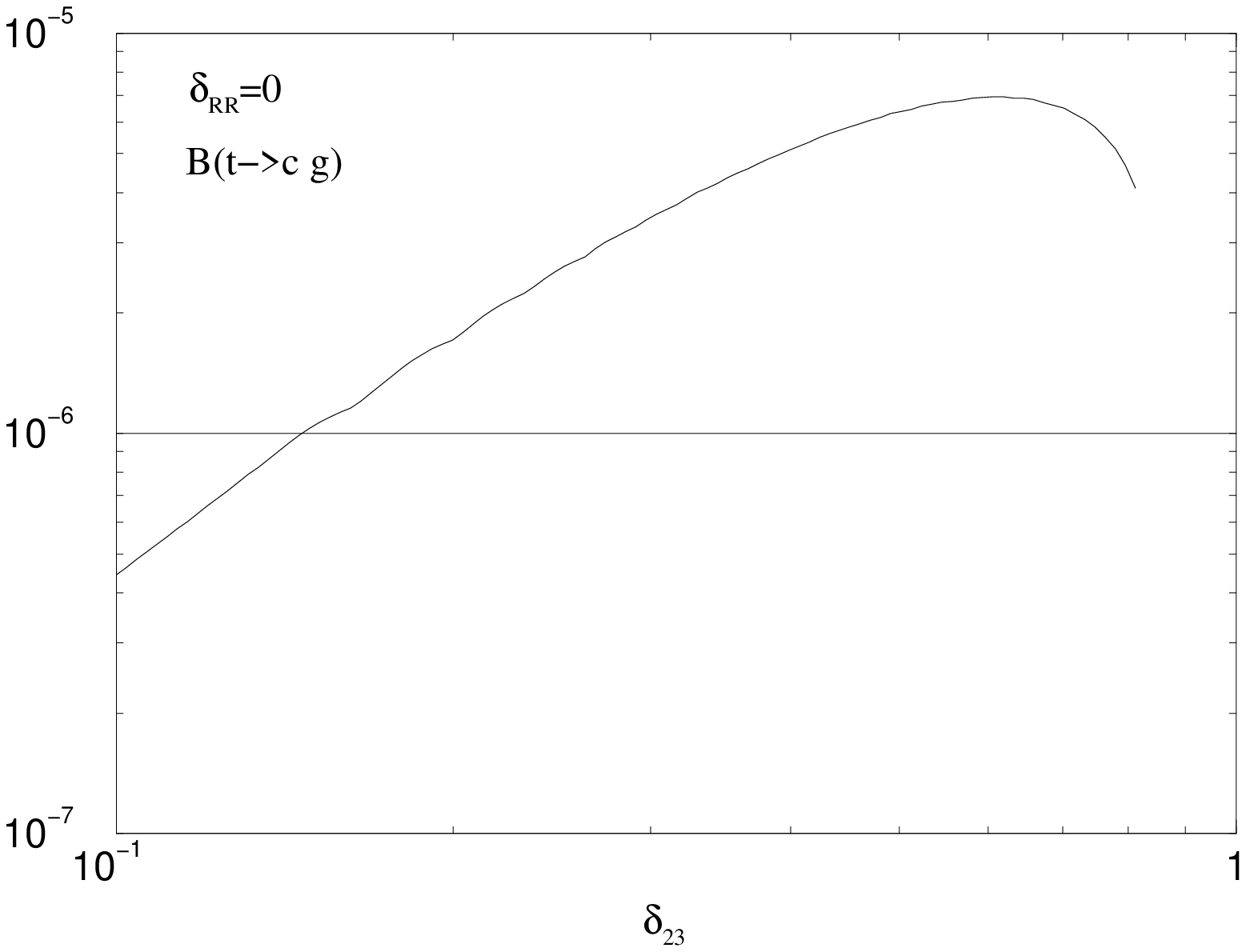}}} \\ 
\multicolumn{2}{c}{(c)}
\end{tabular}
\vspace{.3cm}

{\Large Fig. 10}
\end{center}


\begin{thebibliography}{99}
\bibitem{CGGJS}  J.A. Coarasa, D. Garcia, J. Guasch, R.A. Jim{{\'e}}nez, J.
Sol{{\`a}}, \textit{Eur. Phys. J.} \textbf{C 2} (1998) 373.

\bibitem{MSSM}  H. Nilles, \textit{Phys. Rep.} \textbf{110} (1984) 1; H.
Haber, G. Kane, \textit{Phys.Rep.} \textbf{117} (1985) 75; A.~Lahanas, D.
Nanopoulos, \textit{Phys. Rep.} \textbf{145} (1987) 1; See also the
exhaustive reprint collection \textit{Supersymmetry} (2 vols.), ed. S.
Ferrara (North Holland/World Scientific 1987).

\bibitem{CGHS}  J.A. Coarasa, J. Guasch, W. Hollik, J. Sol{{\`a}}%
, \textit{Phys. Lett. } \textbf{B 442} (1998) 326.

\bibitem{GSPL1}  J. Guasch, J. Sol{\`a}, \textit{Phys. Lett. }\textbf{B 416}
(1998) 353.

\bibitem{Fermilab}  J. Sol{\`a}, talk at the \textit{Physics at Run II
Supersymmetry/Higgs Workshop}, Fermilab, November 19-21, 1998; J.A. Coarasa,
J. Guasch, J. Sol{\`a}, UAB-FT preprint in preparation.

\bibitem{GJHS}  D.~Garcia, R.~A. Jim{\'e}nez, J.~Sol{\`a} and W.~Hollik, 
\textit{Nucl. Phys.} \textbf{B 427} (1994) 53; A.~Dabelstein, W.~Hollik,
C.~J{\"u}nger, R.~A. Jim{\'e}nez and J.~Sol{\`a}, \textit{Nucl. Phys.} 
\textbf{B 454} (1995) 75.

\bibitem{Gianotti}  F.~Gianotti, \textit{Precision Measurements at the LHC},
proc. of the \textit{IVth International Symposium on Radiative Corrections
(RADCOR 98)}, Barcelona, 8-12 September 1998 (World Scientific 1999, to
appear) ed.~J.~Sol{{\`a}}.

\bibitem{Miller}  D.J.~Miller, \textit{Precision studies at a future linear
collider}, [hep-ex/9901039] \newblock proc. of the \textit{IVth International
Symposium on Radiative Corrections (RADCOR 98),} Barcelona, 8-12 September
1998 (World Scientific 1999, to appear) ed.~J.~Sol{{\`a}}; \ See also: 
\textit{2nd Joint ECFA/DESY Study on Physics and Detectors for a Linear
Electron-Positron Collider}, \newblock April 1998--March 1999, Orsay, Lund,
Frascati and Oxford,
\texttt{http://www.desy.de/conferences/ecfa-desy-lc98.html}, papers from the
study to be published in DESY/ECFA report 123F.

\bibitem{Mele}  B.~Mele, S.~Petrarca and A.~Soddu, \textit{Phys. Lett.} 
  \textbf{B 435} (1998) 401. 


\bibitem{Erratum}  G. Eilam, J.L. Hewett, A. Soni, 
  \textit{Erratum: Phys. Rev.} \textbf{D 59} (1998) 039901. 

\bibitem{GEilam}  G. Eilam, J.L. Hewett, A. Soni, \textit{Phys. Rev.} 
  \textbf{D 44} (1991) 1473. 

\bibitem{FCNCSM}  W.S. Hou, \textit{Phys. Lett. }\textbf{B 296} (1992) 179;
  K. Agashe, M. Graesser, \textit{Phys. Rev. }\textbf{D 54} (1996) 4445; M.
  Hosch, K. Whisnant, B.L. Young, \textit{Phys. Rev. }\textbf{D 56}
  (1997) 5725. 

\bibitem{FCNC2HDM} 
H.~Fritzsch, \textit{Phys. Lett. }\textbf{B 224} (1989) 423; 
B.~Grzadkowski, J.F.~Gunion and P.~Krawczyk, \textit{Phys. Lett. }\textbf{B 268} (1991) 106; 
N.G.~Deshpande, B.~Margolis and H.~Trottier, \textit{Phys. Rev. }\textbf{D 45} (1992) 178; 
M.~Luke and M.~Savage, \textit{Phys. Lett. }\textbf{B 307} (1993) 387; 
D.~Atwood, L.~Reina and A.~Soni, \textit{Phys. Rev. }\textbf{D 55} (1997)
3156. 


\bibitem{Yang}  C.S. Li, R.J. Oakes, J.M. Yang, \textit{Phys. Rev. } 
  \textbf{D 49} (1994) 293, \textit{Erratum: ibid. }\textbf{D 56} (1997) 3156.

\bibitem{Koenig}  G. Couture, C. Hamzaoui, H. K{\"o}nig, \textit{Phys. Rev. }
  \textbf{D 52} (1995) 1713; G.~Couture, M.~Frank, H. K{\"o}nig, \textit{Phys.
    Rev. }\textbf{D 56} (1997) 4213.

\bibitem{Lopez}  J.L.~Lopez, D.V. Nanopoulos, R. Rangarajan, \textit{Phys.
    Rev. }\textbf{D 56} (1997) 3100.

\bibitem{Divitiis}  G.M. de Divitiis, R.~Petronzio, L. Silverstini, \textit{%
Nucl. Phys. } \textbf{B 504} (1997) 45. 

\bibitem{FCNCRp}
J.M.~Yang, B.~Young and X.~Zhang,
\textit{Phys. Rev. }{\bf D58} (1998) 055001;
S.~Bar-Shalom, G.~Eilam and A.~Soni, [hep-ph/9812518]
                (\textit{Phys. Rev. }\textbf{D} to appear).

\bibitem{Hunter}  J.F. Gunion, H.E. Haber, G.L. Kane, S. Dawson,\thinspace\ 
\textit{The Higgs Hunters' Guide} (Addison-Wesley, Menlo-Park, 1990).

\bibitem{Higgsloop1}  J. Ellis, G.~Ridolfi, F. Zwirner, \textit{Phys. Lett.}
  \textbf{B 262} (1991) 477; A. Brignole, J. Ellis, G.~Ridolfi, 
F. Zwirner, \textit{Phys. Lett.} \textbf{B 271} (1991) 123; H.E.
Haber, R.~Hempfling, \textit{Phys. Rev. Lett.} \textbf{66} (1991)
1815; \textit{Phys. Rev. }\textbf{D 48} (1993) 4280; M. Carena, J. Espinosa,
M. Quir{\'o}s, C. Wagner,  \textit{Phys. Lett.} \textbf{B 355} (1995) 209.

\bibitem{Dabelstein}  P.~Chankowski, S.~Pokorski and J.~Rosiek,
  \textit{Nucl. Phys.} \textbf{B 423} (1994) 437; A.~Dabelstein,
  \textit{Z. Phys.} \textbf{C 67 }(1995) 495; A.~Dabelstein,
  \textit{Nucl. Phys.} \textbf{B 456} (1995) 25.

\bibitem{Higgsloop2}  M. Carena, M. Quir{\'o}s, C.
  Wagner, \textit{Nucl. Phys.} \textbf{B 461} (1996) 407;  H. Haber,
  R.~Hempfling, A. Hoang, \textit{Z. Phys}. \textbf{C 75} (1997) 539;  
  S. Heinemeyer, W. Hollik, G.~Weiglein, \textit{Phys. Lett.} \textbf{B 440}
  (1998) 296.

\bibitem{HollWeig}  Talks by W. Hollik and G. Weiglein [hep-ph/9901317] in: 
  \textit{IVth International Symposium on Radiative Corrections (RADCOR 98),}
  Barcelona, 8-12 September 1998 (World Scientific 1999, to appear)
  ed.~J.~Sol{\`a}. 


\bibitem{tch}  J.M. Yang, C.S. Li, \textit{Phys. Rev. }\textbf{D 49} (1994)
3412, \textit{Erratum: ibid. }\textbf{D 51} (1995) 3974.

\bibitem{Prelim}  {J.~Guasch, in: \textit{Quantum Effects in the MSSM}
    (World Scientific 1998,  p. 256) (ISBN 981-02-3450-3) ed.~J.~Sol{\`a};  
    J. Guasch, talk at the  \textit{2nd Joint ECFA/DESY Study on Physics and
      Detectors for a Linear Electron-Positron Collider}, 20th-23th March 1999,
    Oxford, U.K.; J. Sol{\`a}, talk at the 
    \textit{Workshop on the LC Collider}, Sitges, Spain, April 28th-May 5th
    1999.} 

\bibitem{GS1}  J. Guasch, J. Sol{\`a}, \textit{Z. Phys. }\textbf{C 74}
(1997) 337.

\bibitem{duncan}  M.J. Duncan, \textit{Nucl. Phys. }\textbf{B 221} (1983)
285;  \textit{Phys. Rev. }\textbf{D 31} (1985) 1139.

\bibitem{gabbiani}  F. Gabbiani, E. Gabrielli, A. Masiero, L. Silverstrini, 
\textit{Nucl. Phys. }\textbf{B 477} (1996) 321.

\bibitem{Pokorski}  M. Misiak, S. Pokorski, J. Rosiek, \textit{Supersymmetry
and FCNC Effects}, in: ``Heavy Flavours II'', eds. A.J.~Buras, M. Lindner,
Advanced Series on directions in High Energy Physics, World Scientific 1998.


\bibitem{SUGRA}  A.~Bouquet, J.~Kaplan and C.A.~Savoy, \textit{Phys. Lett. }{\bf B 148}
 (1984) 69; S. Bertolini, F. Borzumati, A. Masiero, G. Ridolfi,
  \textit{Nucl. Phys. }\textbf{B 353} (1991) 591; C. Kolda, L. Roszkowski,
  J.D. Wells, G.L. Kane, \textit{Phys. Rev. }\textbf{D 50} (1994) 3498. 


\bibitem{UAB97}  See the contributions by P. Giacomelli, M. Williams, A.
  Dominguez and R. Keranen,\thinspace\ in: \textit{Quantum Effects in the MSSM}
  (World Scientific 1998) (ISBN 981-02-3450-3) ed.~J.~Sol{\`a}.

\bibitem{JGI}  J.~Guasch, in: \textit{Quantum Effects in the MSSM} (World
Scientific 1998, p. 256) (ISBN 981-02-3450-3) ed.~J.~Sol{\`a}.



\bibitem{limits}  R. Frey \textit{et al.}, \textit{Top Quark Physics: Future
Measurements}, preprint FERMILAB-CONF-97-085, April 1997 [hep-ph/9704243].

\bibitem{CJS}  J.~A. Coarasa, R.~A. Jim{\'e}nez and J.~Sol{{\`a}},
  \textit{Phys. Lett.} \textbf{B 389} (1996) 312.  


\bibitem{Djouadi} A. Djouadi, in: \textit{Quantum Effects in the
MSSM} (World Scientific 1998, p. 197) (ISBN 981-02-3450-3) ed.~J.~Sol{\`a};
M. Spira, \textit{Fortsch. Phys.} 46 (1998) 203.


\end{thebibliography}
\end{document}